\documentclass[a4paper,fleqn,usenatbib]{mnras}

\usepackage{newtxtext,newtxmath}

\usepackage[T1]{fontenc}
\usepackage{ae,aecompl}

\usepackage{graphicx}	% Including figure files
\usepackage{epstopdf}
\usepackage{amsmath}	% Advanced maths commands
\def\VEV#1{\left\langle #1\right\rangle} % This is \VEV{x} => <x>

\usepackage{etoolbox}

\title[Quasi-equilibrium models of star-forming disc galaxies]{Quasi-equilibrium models of high-redshift disc galaxy evolution}

\author[Furlanetto]{
Steven R.~Furlanetto$^{1}$\thanks{E-mail: sfurlane@astro.ucla.edu} \\
$^{1}$Department of Physics \& Astronomy, University of California, Los Angeles, Los Angeles, CA 90095, USA\\
}

\date{Accepted XXX. Received YYY; in original form ZZZ}

\pubyear{2015}

\begin{document}
\label{firstpage}
\pagerange{\pageref{firstpage}--\pageref{lastpage}}
\maketitle

\begin{abstract}
In recent years, simple models of galaxy formation have been shown to provide reasonably good matches to available data on high-redshift luminosity functions. However, these prescriptions are primarily phenomenological, with only crude connections to the physics of galaxy evolution. Here we introduce a set of galaxy models that are based on a simple physical framework but incorporate more sophisticated models of feedback, star formation, and other processes. We apply these models to the high-redshift regime, showing that most of the generic predictions of the simplest models remain valid. In particular, the stellar mass--halo mass relation depends almost entirely on the physics of feedback (and is thus independent of the details of small-scale star formation) and the specific star formation rate is a simple multiple of the cosmological accretion rate. We also show that, in contrast, the galaxy's gas mass is sensitive to the physics of star formation, although the inclusion of feedback-driven star formation laws significantly changes the naive expectations. While these models are far from detailed enough to describe every aspect of galaxy formation, they inform our understanding of galaxy formation by illustrating several generic aspects of that process, and they provide a physically-grounded basis for extrapolating predictions to faint galaxies and high redshifts currently out of reach of observations. If observations show violations from these simple trends, they would indicate new physics occurring inside the earliest generations of galaxies.
\end{abstract}


%%%%%%%%%%%%%%%%%%%%%%%%%%%%%%%%%%%%%%%%%%%%%%%%%%

%%%%%%%%%%%%%%%%% BODY OF PAPER %%%%%%%%%%%%%%%%%%

\section{Introduction} \label{intro}

Galaxy evolution is one of the most fundamental aspects of astrophysics, but it remains mysterious even today thanks to a number of complex processes including star formation, turbulent feedback, chemical evolution, and disc dynamics. These and other physical mechanisms, many of them only crudely understood, interact to determine each system's evolution. Over and above these internal processes, galaxies are also influenced by their environments: nearby systems interact with each other, and radiation backgrounds can affect the state of the gas inside galaxies, especially at early times. The complex interplay of all these processes presents a challenging physics problem, and for many years galaxy evolution was approached almost exclusively through numerical simulations and semi-analytic models. Many of these have focused on the first billion years of the Universe's history, including \citet{dayal13, mutch16, yung19, vogelsberger20, hutter20}.

Meanwhile, over the past few decades, a number of large surveys have measured the properties of galaxy populations from the present day, back through the ``cosmic noon" of star formation $z \sim 2$--3, and even during the first billion years of galaxy formation ($z \ga 6$). These studies have shown that most galaxies are part of well-defined populations that obey apparently simple scaling relations, such as the fundamental plane, the Tully-Fisher relation, and either the star-forming main sequence or the red sequence. The evident simplicity of these relations suggests the existence of fundamental and generic mechanisms at the heart of galaxy formation.

In the past decade, several simple, analytic models have emerged to try to explain that simplicity. ``Semi-empirical models" use a small number of relations, calibrated to observations and sometimes more complicated numerical simulations, to describe the evolving galaxy population (e.g., \citealt{behroozi13, tacchella13, mason15, sun16, mirocha17, tacchella18, behroozi19}). Such models have shown, for example, that the halo accretion rate and the star formation efficiency  follow simple trends. Although the inferred relations match qualitative expectations of galaxy formation, these models do not try to identify a specific physical explanation for them.  

A complementary set of simple analytic galaxy evolution models have attempted to elucidate these trends \citep{bouche10, dave12, dekel13, lilly13, furl17-gal, mirocha20}. For example, the ``minimal bathtub model" \citep{dekel13} assumes that galaxies are machines that accrete gas onto an interstellar medium (ISM), from which stars form through some imposed prescription and feedback from those stars ejects a fraction of the gas. The bathtub model suggests some universal aspects of galaxy formation (which we will examine in section \ref{min-bath-intro}), such as a tight link between the specific star formation rate and the cosmological accretion rate. Several other models have shown that the balance between star formation and feedback provides a reasonable fit to luminosity functions at very high redshifts \citep{trenti10, tacchella13, tacchella18, mason15, furl17-gal}. (The high-redshift regime is a useful test of such models, because many of the most poorly-understood galaxy evolution processes, such as galaxy quenching, are largely irrelevant in this regime.) 

However, such models are only a small step away from semi-empirical treatments, because they take such a simplistic view of the underlying galaxy formation processes. It is thus not clear whether their conclusions remain valid in more sophisticated (and, one hopes, realistic) models. In particular, they cannot leverage the improvements made over the past several years in our understanding of star formation and feedback. Over this time, analytic descriptions of these processes have improved tremendously, often informed by the results of detailed numerical simulations. For example, \citet[hereafter FQH13]{faucher13} showed how the assumption that stellar feedback provides turbulent support to galaxy discs can fix the star formation rate. In contrast, \citet[hereafter K18]{krumholz18} showed how other support mechanisms (principally radial transport), as well as the chemistry of molecular hydrogen, affect the star formation law. \citet[hereafter S18]{semenov18} provided a third approach, in which star formation is regulated by the cycling of gas between a star-forming phase and the general ISM. Meanwhile, \citet{thompson16} and \citet{hayward17} have shown how feedback can be modeled efficiently in an inhomogeneous galaxy disc. While there remains controversy over the connections between these small-scale processes and the large-scale galactic environment, these approaches offer a much closer window into the fundamental physics of star formation than the minimalist models. 

In this paper, we introduce a new set of analytic galaxy evolution models that use these state-of-the-art analytic prescriptions for star formation and feedback to track galaxy growth over cosmological timescales, providing a flexible framework to contrast the predictions of different pictures. We build these prescriptions on the foundation of the simple minimalist bathtub model in order to test how its generic predictions fare when more physical processes are included. We intentionally include a broad range of star formation models in order to test the importance of assumptions underlying those processes. Including more sophisticated treatments also allows us to examine additional galaxy observables (such as the gas mass, the circumgalactic medium, and chemistry), permitting us to examine whether population-level measurements can distinguish different models of star formation and feedback. 

While all of these prescriptions are still simplistic compared to full semi-analytic models and numerical simulations, the flexibility of our treatment allows us to vary the assumptions behind star formation and feedback quite dramatically and identify trends that are robust with respect to these basic  processes. Such results help us to understand how we might extrapolate our understanding of galaxy physics to unknown populations (including both faint and high-$z$ galaxies). They also provide a ``null test" to help identify flaws in our basic understanding of galaxies. For example, we shall see that the stellar mass evolution in all of our models is nearly independent of the star formation law but depends strongly on the feedback prescription. If observations find qualitatively different stellar mass evolution than our models, this suggests that we should focus on understanding feedback to improve the models.

Of course, these models are by no means a first-principles description of galaxy evolution. Of particular importance, we assume that galaxies form discs that support themselves in a quasi-equilibrium state over cosmological timescales, as also assumed by the star formation prescriptions we use. This is likely \emph{not} true at very early times: the halo dynamical time is $\sim 1/\sqrt{G \bar{\rho}_h} \sim t_H/\sqrt{18 \pi^2}$, where $\bar{\rho}_h \sim 18 \pi^2 \bar{\rho}_m$, $\bar{\rho}_h$ is the mean halo density, $\bar{\rho}_m$ is the mean matter density in the Universe, and $t_H$ is the Hubble time. At these early times, this dynamical time can be comparable to or even smaller than the timescale for massive stars to evolve, so that it will be difficult for stellar feedback and star formation to come into equilibrium \citep{faucher18, orr19}. This naturally explains the burstiness found in simulated galaxies during this era (e.g., \citealt{teyssier13, hopkins14, agertz15, kimm15}), an aspect that we will not try to model. Our model is probably more applicable to later times during this era, and to relatively large galaxies, where the quasi-equilibrium approach is most likely to be valid. That said, the extrapolation to earlier times does help to define a baseline against which non-equilibrium effects can be compared.

We also note that the equilibrium assumption can be problematic at later times as well. For example, many local dwarf galaxies appear to have gas depletion times comparable to or even longer than the Hubble time \citep{cannon15, jameson16}. In such a case, star formation does not proceed rapidly enough to bring the galaxy into the equilibrium state (unless the star formation is accompanied by strong feedback driving rapid mass loss, but that does not appear to be the case in these systems), so that more sophisticated prescriptions are necessary (e.g., \citealt{forbes14b}). We will not apply our formalism to dwarfs at low redshifts, but we caution the reader that similar effects may apply to dwarf galaxies at early times.

In section \ref{dmhaloes}, we describe our assumptions about dark matter halo growth. In section \ref{min-bath-intro}, we review the simplest ``bathtub" galaxy formation models, including how cosmological evolution affects their predictions. Then, in section \ref{power-law-model} we show how some simple physically-motivated prescriptions for star formation and feedback affect the bathtub model predictions. In section \ref{met-rec}, we supplement the model with descriptions of gas recycling and chemical evolution. In section \ref{feedback-models}, we introduce some state-of-the-art models that motivate the simple prescriptions of the previous section, while in section \ref{sf-phase} we describe models that include a star-forming gas-phase. Next, in section \ref{gal-zevol}, we examine how individual galaxies evolve over time in these models, and in section \ref{mass-trends} we examine their consequences for galaxy populations and scaling relations. Finally, we conclude in section \ref{disc}.

The numerical calculations here assume $\Omega_m = 0.308$, $\Omega_\Lambda = 0.692$, $\Omega_b = 0.0484$, $h=0.678$, $\sigma_8=0.815$, and $n_s=0.968$, consistent with the recent results of \citet{planck18}. Unless specified otherwise, all distances quoted herein are in comoving units.

\section{The cosmological context: dark matter haloes} \label{dmhaloes}

\subsection{Halo growth} \label{halo-grow}

We assume that galaxies inhabit dark matter haloes and let $n_h(m,z)\, dm$ be the comoving number density of dark matter haloes with masses in the range $(m, m+dm)$ at a redshift $z$.  We write this as
\begin{equation}
n_h(m,z) = f(\sigma) { \frac{\bar{\rho}}{ m }} {d \ln \frac{(1/\sigma)}{dm}},
\label{eq:nhm}
\end{equation}
where $\bar{\rho}$ is the comoving matter density, $\sigma(m,z)$ is the rms fluctuation of the linear density field, smoothed on a scale $m$, and $f(\sigma)$ is a dimensionless function. We take $f(\sigma)$ from a fit to recent high-$z$ cosmological simulations \citep{trac15}. This version differs from other common implementations \citep{sheth02, tinker08-hmf} by $\sim 20$\% for halo masses $\ga 10^{10} \, M_\odot$, but because we only use the halo abundances to derive accretion rates in this paper, such uncertainties do not affect our results significantly.

We also assume that haloes begin forming stars above a specified minimum mass  $m_{\rm min}$, which we take to be equivalent to a virial temperature of $10^4$~K, where hydrogen lines allow gas to cool to high densities \citep{loeb13}. 

Our model follows haloes as they grow galaxies, so we require the halo accretion rates. In the spirit of constructing a simple model, we will assume that this accretion occurs smoothly and steadily, varying significantly only on cosmological timescales. Simulations have commonly measured relations similar to \citep{mcbride09, fakhouri10, goerdt15, trac15}
\begin{equation}
\dot{m}_h = A m_h^{\mu_h} (1+z)^\beta,
\label{eq:sim-gen}
\end{equation}
where $A \sim 0.03$~Gyr$^{-1}$ is a normalization constant, $m_h$ is the halo mass, $\mu \ga 1$ in the simulations,\footnote{For example, \citet{mcbride09} find $\mu_h=1.127$ at moderate redshifts, while \citet{trac15} find $\mu_h =1.06$ at $z \sim 6$--10 and $10^8 \la (m_h/M_\odot) \la 10^{13}$.} and $\beta \approx 5/2$ in the matter era. \citet{dekel13} (see also \citealt{neistein06, neistein08}) argued that this form can be understood through the extended Press-Schechter approach \citep{lacey93}, in which $\mu_h$ comes from the shape of the matter power spectrum, while $\beta \approx 5/2$ emerges from the redshift dependence of the halo mass function.

However, this relation has not been tested at very high redshifts or at very small masses, which are of particular interest to us. We therefore use a slightly more sophisticated prescription for the accretion rate in our full model, though we will use equation~(\ref{eq:sim-gen}) to develop intuition. The model is described in detail in \citet{furl17-gal}, but in short we assume that haloes maintain their number density as they evolve, in a similar fashion to abundance matching \citep{vale04, vandokkum10}. We demand that at any given pair of redshifts $z_1$ and $z_2$ a halo has masses $m_1$ and $m_2$ such that
\begin{equation}
\int_{m_1}^\infty dm \, n_h(m,z_1) = \int_{m_2}^\infty dm \, n_h(m,z_2).
\label{eq:halo-mod-acc}
\end{equation}
The accretion rate $\dot{m}(m,z)$ follows by assuming that this is true for all redshifts and all halo masses.   \citet{furl17-gal} showed that this model agrees reasonably well with the simulation rates where the comparison can be made but departs at high redshifts and small masses. Like equation~(\ref{eq:sim-gen}), it leads to roughly exponential halo growth, $m_h(z) \propto e^{-\alpha z}$. 

For simplicity, we will ignore scatter in this relation: though mergers are an important part of halo growth, even at moderate redshifts the majority of matter is acquired through smooth ongoing accretion \citep{goerdt15}. This assumption allows us to build our model as a simple set of differential equations.

\subsection{Halo properties} \label{halo-prop}

We assume that galaxy discs have radii proportional to their halo virial radii, $r_{\rm vir} \propto m_h^{1/3} (1+z)^{-1}$, with a half-mass radius $r_d = A_r r_{\rm vir}$, where $A_r $ is a constant determined, in the standard disc formation model, by the spins imparted to the halo through large-scale torques \citep{mo98}. We take $A_r = 0.02$ as a fiducial value, slightly larger than expected from \citet{kravtsov13} and \citet{somerville18}. Thus
\begin{equation}
r_d = 0.28 \left( {A_r \over 0.02} \right) \left( {m_h \over 10^{11} \ M_\odot} \right)^{1/3} \left( {1 + z \over 7} \right)^{-1} \ {\rm kpc}.
\label{eq:disk-radius}
\end{equation}
We also assume for simplicity that disks have uniform surface density and take $m_g = 2 \pi \Sigma_g r_d^2$. 

Next, we require a relationship between the rotational velocity in the disc and the host halo. We assume the discs are embedded in cosmological halos following the \citet{navarro97} profile, whose key parameter is the halo concentration, which determines the halo's scale radius, $r_s = r_{\rm vir}/c_h$. We use 
\begin{equation}
c_h = {15 \over 1+z} \left( {m_h \over 10^{12} \, M_\odot} \right)^{-0.2},
\end{equation}
which was proposed by \citet{faucher18} as a compromise between earlier results \citep{seljak00,bullock01}. In such halos, the maximum orbital velocity is
\begin{equation}
v_{\rm max} \approx 0.46 \left( {\ln (1 + c_h) \over c_h} - {1 \over 1 + c_h} \right)^{-1/2} v_c,
\label{eq:vmax}
\end{equation}
where $v_c$ is the halo circular velocity. We then assume that, if dark matter were the only source of gravity, the orbital velocity within the galaxy disc would be a constant multiple of this, $v_{\rm orb,d} = A_{\rm orb} v_{\rm max}$. We take $A_{\rm orb} = 0.9$, which \citet{faucher18} argued matches the Tully-Fisher relation at $z \la 2$. 

We estimate the contribution of the baryonic component's gravity to the rotational velocity by assuming that it is distributed spherically (or, equivalently for our purposes, has a Mestel disk profile). Then the contribution of baryons to the orbital velocity is $v_{\rm orb,b}^2 = \sqrt{G m_b/r_d}$, where we take $m_b$ to include both stars and gas in the disk and assume that half of the total baryonic mass in the disk is inside the scale radius. However, with our assumptions the baryonic contribution is almost always subdominant. 

Finally, the orbital timescale is $t_{\rm orb} \approx 2 \pi r_d/v_{\rm orb}$. For context, it is useful to note that the orbital times in these galaxies are very short indeed. A useful approximation can be attained by assuming an isothermal dark matter profile, with no baryons, and evaluating the orbital timescale $t_{\rm orb}$ at the characteristic disc radius $r_d$. Then, 
\begin{equation}
t_{\rm orb} \sim 18 \left( {A_r \over 0.02} \right) \left( {7 \over 1+z} \right)^{3/2} \ {\rm Myr}.
\label{eq:torb}
\end{equation}

\section{The bathtub model} \label{min-bath-intro}

We now describe a simple model of galaxy formation within these dark matter haloes. We will construct the model in stages so as to emphasize the relevant physics and its uncertainties. We begin with a model similar to the ``bathtub" model \citep[see also the similar ``regulator" model of \citealt{lilly13}]{bouche10, dave12, dekel14}. First, we assume that baryonic matter accretes onto a halo with a rate $\dot{m}_{c,g} = (\Omega_b/\Omega_m) \dot{m}_h$ fixed by cosmology. This gas builds a reservoir (the galaxy's diffuse ISM) of mass $m_g$ from which stars will form.\footnote{A more realistic treatment allows a fraction $f_{c,*}$ of this material to arrive in the form of stars, with the remainder being gas; see \citet{dekel14}. We simplify our picture by focusing on the buildup of a monolithic galaxy.} 

Stars form within the ISM at a rate $\dot{m}_{*}$. For now, we write this rate as $\dot{m}_{*} = m_g/t_{\rm sf}$, where $t_{\rm sf}$ is a parameter to be determined that may depend on the halo properties. We also assume that feedback from the stars returns a fraction $(1-\mu)$ of this material to the star-forming phase of the ISM on very short timescales. The return fraction obviously depends on the definition of ``very short timescales" and the properties of the stellar population, but for simplicity we take a constant $\mu=0.25$, which roughly matches the return fraction for a Chabrier IMF after 100~Myr (see Fig.~11 of \citealt{benson10}). The more detailed model of \citet{tacchella18} find that this estimate is reasonable at $z \la 6$ but overestimates the return fraction at earlier times. We do not consider time evolution in order to keep our model as simple as possible. Feedback driven by star formation also ejects mass from the diffuse gas at a rate $\eta \dot{m}_{*}$, where in general the mass-loading parameter $\eta$ is also a function of the galaxy's properties. Then we can write
\begin{eqnarray}
\dot{m}_g & = & \dot{m}_{c,g} - (\mu+ \eta) \dot{m}_{*}, \label{eq:g-gas-bt} \\
\dot{m}_{*} & = & m_g/t_{\rm sf}. \label{eq:g-star-bt} 
\end{eqnarray}
Note that, because only a fraction $\mu$ of the mass forming stars remains inside those objects, the stellar mass at later times is $\mu$ multiplied by the integral of the star formation rate.

To develop intuition, for now we will forego our abundance-matching accretion rate prescription and instead use the approximate form for the accretion rate in equation~(\ref{eq:sim-gen}), setting $\mu_h=1$. We will consider the evolution of a galaxy inside a halo that begins to form stars when it has a mass $m_0$ at $z_0$. Then,
\begin{equation}
m_h(z) = m_{0} e^{-\alpha_h (z - z_0)},
\label{eq:halo-growth}
\end{equation}
for some constant $\alpha_h$. Excursion set estimates suggest $\alpha \approx 0.79$ at moderate redshifts \citep{neistein08}, but our abundance matching approach finds somewhat larger values at early times. We will also assume for simplicity  that $\eta \gg 1 > \mu$, so that $\mu + \eta \approx \eta$.

\subsection{The bathtub model in its simplest incarnation} 

\citet{dekel14} made simple choices for the input parameters, taking $\eta$ as a constant (over both mass and redshift) and assuming the star formation time to be proportional to the galaxy orbital timescale, $t_{\rm sf} = t_{\rm orb}/\epsilon$ with $\epsilon$ also a constant. In this picture, a constant fraction of the galaxy's gas is cycled into stars over each orbital period.  In that case, if the accretion rate and star formation timescale are treated as constants, equations~(\ref{eq:g-gas-bt}) and (\ref{eq:g-star-bt}) have analytic solutions, in which the gas mass approaches a constant value, $m_g(t) = \dot{m}_{c,g} \tau (1 - e^{-t/\tau})$, where $\tau \approx t_{\rm sf}/\eta$. If the accretion rate changes on timescales much larger than $\tau$, this reduces to $m_g(t) \approx  \dot{m}_{c,g} \tau$, where now both factors on the right-hand side are (slowly-varying) functions of cosmic time. 

In this limiting case, \citet{dekel14} write $m_g$ as a fraction of $m_a = (\Omega_b/\Omega_m) m_h$, the total baryonic mass that has ever accreted onto the halo. We will refer to this as the \emph{gas retention fraction} $X_g \equiv m_g/m_a$.  Then,
\begin{equation}
X_g \approx {t_{\rm sf} \over \eta t_{\rm acc}} = {1 \over \epsilon \eta} {t_{\rm orb} \over t_{\rm acc}} , 
\label{eq:xg-min}
\end{equation}
where the accretion time $t_{\rm acc} = m_a/\dot{m}_{c,g}$. Thus $X_g \propto (1+z)/\epsilon \eta$ in this simple model: the relative size of the galaxy's gas reservoir decreases slowly with time.

\citet{dekel14} identified four basic implications of this simple picture, which also occur in similar models \citep{lilly13}: \emph{(i)} Because $m_g(t) \approx  \dot{m}_{c,g} \tau$, the star formation rate is $\dot{m}_* = \dot{m}_{c,g}/\eta$. In this limit, the ISM reaches a steady state in which gas accretion is balanced by star formation and its associated feedback and is surprisingly \emph{independent} of the star formation efficiency $\epsilon$. \emph{(ii)} On the other hand, $X_g$ -- and hence the gas mass itself -- is inversely proportional to $\epsilon$, because a lower star formation efficiency requires the ISM to have more fuel in order to maintain the equilibrium state. \emph{(iii)} Like the dark matter halo itself, the stellar content of halos grow exponentially with redshift, because $\dot{m}_* \propto \dot{m}_{c,g}$. \emph{(iv)} Finally, in this equilibrium state the star formation rate is directly proportional to the overall accretion rate. Thus the specific star formation rate (SSFR) only depends on the accretion timescale $t_{\rm acc}$ -- all the other parameters cancel in the ratio $\dot{m}_* / m_*$. Measurements of the SSFR will only probe the physics of star formation to the extent that the minimal bathtub model's assumptions are violated.

\subsection{The bathtub model with cosmological evolution} 

We now incorporate cosmological evolution explicitly into the bathtub model. To do so, it is convenient to rewrite the equations in terms of redshift, so we will define primes to be derivatives with respect to $z$. We will also define $\tilde{m} \equiv m/m_0$ and write 
\begin{equation}
t_{\rm orb} \equiv t_{\rm orb,0} \left( {1 + z_0 \over 1 + z} \right)^{3/2} \equiv {A_{\rm dyn} \over \sqrt{\Delta_{\rm vir}}} H(z)^{-1},
\end{equation}
where $\Delta_{\rm vir}$ is the virial overdensity and $A_{\rm dyn}$ is a correction of order unity. (The redshift dependence here is approximate, ignoring changes to the halo and disc structure.)

Then the relevant equations become
\begin{eqnarray}
{\tilde{m}_h' \over \tilde{m}_h} & = &  - | \tilde{m}_0'| \label{eq:bathtub-dm} \\
{\tilde{m}_g' \over \tilde{m}_g } & = & - | \tilde{m}_0'| \left[ X_g^{-1} - {\epsilon (\mu  + \eta) \over | \tilde{m}_0'| C_{\rm orb}} \left( {1 + z \over 1 + z_0} \right) \right], \label{eq:bathtub-mg-analytic} \\
\tilde{m}_*'  & = & -\mu \tilde{m}_g  \left[ {\epsilon \over C_{\rm orb}} \left( {1 + z \over 1 + z_0} \right) \right], \label{eq:bathtub-sfr-analytic} \\
X_g' & = & X_g \left( {\tilde{m}_g' \over \tilde{m}_g} - {\tilde{m}_h' \over \tilde{m}_h} \right), \label{eq:bathtub-xg-analytic}
\end{eqnarray}
where $\tilde{m}_0' \equiv - \dot{m}_0/m_0 \times [H(z_0) (1+z_0)]^{-1}$ and $C_{\rm orb} = (1+z_0) A_{\rm dyn} \Delta_{\rm vir}^{-1/2}$ are both constants for a given halo. Here $\Delta_{\rm vir} = 18 \pi^2$ is the virial overdensity of a collapsed halo. The last of these equations is not independent of the others, but it will be useful for providing intuition.

In the last section, we found that $X_g$ varies relatively slowly with time -- unlike the halo and stellar masses, which grow exponentially.  We also expect that $X_g \ll 1$, because much of the gas will be expelled through feedback and/or turned into stars. Then, if $X_g' \approx 0$,  the term in parentheses in equation~(\ref{eq:bathtub-xg-analytic}) must vanish. This factor is determined by a combination of three processes: dark matter accretion, gas accretion, and star formation (plus its associated feedback). But the two accretion terms have very different  effects: the gas component is enhanced by a factor $X_g^{-1} \gg 1$, because the gas disc contains only a small fraction of the halo's baryonic mass. Thus cosmological accretion has a much more significant \emph{relative} effect on the baryonic component. In order to drive $X_g' \approx 0$, we therefore must have the cosmological gas accretion nearly cancelled by the combination of star formation and feedback, which is described by the last term in equation~(\ref{eq:bathtub-mg-analytic}). Examining these terms, we see that we must have $X_g \propto 1/[\epsilon (\mu+\eta)]$ for this cancellation to occur (though there is also some modest redshift dependence). Because $m_g \propto X_g$, the gas reservoir has the same scaling, and we recover essentially the same behavior as in the simpler model, including the parameter dependence.  The key difference is that the gas reservoir grows exponentially with time, so as to keep $X_g$ roughly constant.

\begin{figure}
	\includegraphics[width=\columnwidth]{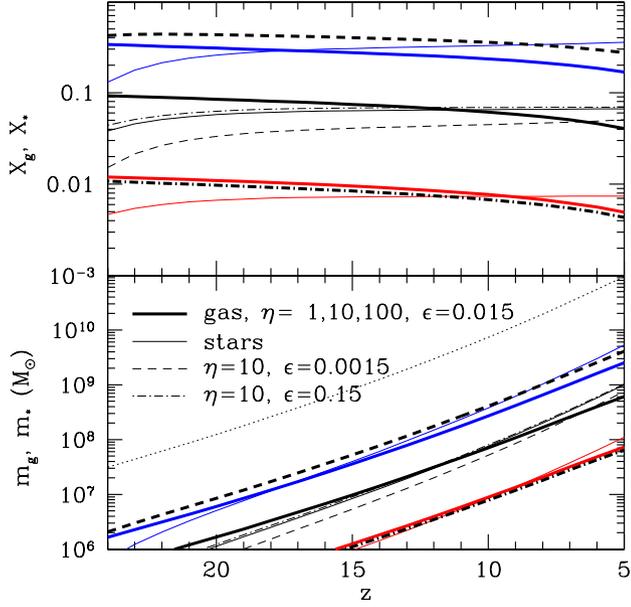} 
    \caption{Solutions to the minimal bathtub model for a halo with $m \approx 10^{11} \, M_\odot$ at $z=5$. The bottom panel shows the halo mass (dotted curve), gas mass (thick curves), and stellar mass (thin curves). The top panel shows the gas and stellar fractions, relative to $(\Omega_b/\Omega_m) m_h$.}
    \label{fig:bathtub}
\end{figure}

Figure~\ref{fig:bathtub} illustrates these results with (exact) solutions to the model. In this, and in our later calculations, we begin tracking a halo when it crosses the threshold for atomic cooling at $z_i$. We initialize our galaxies with simple estimates of the gas and stellar fractions and then solve our evolution equations until they converge to smooth solutions (which usually occurs after $\Delta z \approx 1$). We then re-initialize the halo at $z_i$ using the gas and stellar fractions from the smooth solutions. This procedure to set our initial conditions reduces transient behavior from an initial burst of star formation, which is not captured by the assumptions of our model. (Regardless of our choice of initial conditions, the rapid growth of halos in these regimes renders our assumptions unimportant for $z \la z_i-1$.) 

The bottom panel of Figure~\ref{fig:bathtub} shows the total mass in the gas and stellar phases, while the top panel normalizes those quantities to the baryonic mass associated with the halo (so $X_* = m_*/m_a$). Consider first the solid curves, which vary the feedback parameter $\eta$ across two orders of magnitude. Both the gas and stellar masses are nearly proportional to $(\mu+\eta)$. Meanwhile, the dashed and dot-dashed curves vary $\epsilon$ across two orders of magnitude. While the gas masses vary strongly across these cases, as expected, the stellar masses are mostly independent. The exception is the low efficiency case ($\epsilon=0.0015$), for which the star formation timescale is so long that the initial transient adjustment is quite long. Finally, also note that both $X_g$ and $X_*$ are nearly constant over this entire period -- with the gas fraction decreasing slowly (as more is converted to stars) and the stellar fraction increasing slowly.

%FIGURE: Bathtub models

\section{Power-Law Prescriptions for Star Formation and Feedback} \label{power-law-model}

The key shortcoming of the model so far is that the parameters $\epsilon$ and $\eta$ are constants that must be prescribed by hand. Next, we will show that more flexible prescriptions for these efficiencies do not strongly affect the main conclusions of the model. We begin by describing more general parameterizations (including their motivation) in sections \ref{sfr-sd} and \ref{fg-simple}, and we incorporate them into our model in section \ref{insight-power-law}.

\subsection{The Star Formation Rate--Surface Density Relation} \label{sfr-sd}

Motivated by the observed relations between the surface density of the star formation rate, $\dot{\Sigma}_* $, and the total gas surface density, $\Sigma_g$, we will now introduce star formation prescriptions that depend on $\Sigma_g$. Most famously, \citet{kennicutt98} showed that, when averaged over large regions of galaxies, \begin{equation}
\dot{\Sigma}_* \approx 10^{-5} \left( { \Sigma_g \over M_\odot \, {\rm pc}^{-2}} \right)^{1.4} \ M_\odot \, {\rm pc}^{-2} \, {\rm Myr}^{-1} .
\label{eq:KSlaw}
\end{equation}
Since then, many observations have further explored this Kennicutt-Schmidt law (e.g., \citealt{kennicutt12}), including at smaller scales (e.g., \citealt{kennicutt07}) and higher redshifts (e.g., \citealt{bouche07}), and exploring relations between $\dot{\Sigma}_* $ and the molecular gas content and/or orbital time scale (e.g., \citealt{leroy13, tacconi13}).

To interpret the Kennicutt-Schmidt law, it is convenient to write
\begin{equation}
\dot{\Sigma}_* \equiv \epsilon_{\rm ff}^{\rm gal} {\Sigma_g \over t_{\rm ff}^{\rm disc}},
\label{eq:sfr-breakdown}
\end{equation}
where $\epsilon_{\rm ff}^{\rm gal}$ is the star formation efficiency per disc free-fall time averaged over the entire galaxy and $t_{\rm ff}^{\rm disc}$ is the free-fall time at the mean density of the disc. If gas clouds form stars with a fixed efficiency $\epsilon_{\rm ff}^{\rm gal}$ per free fall time $t_{\rm ff}^{\rm disc}$, if the disc dominates the local density (and hence the free-fall time), and if $\Sigma_g \propto \rho_g$, the local three-dimensional gas density, we would have $\dot{\Sigma}_* \propto \Sigma_g^{3/2}$, close to the observed relation. Many local observations show that $\epsilon_{\rm ff}^{\rm gal} \approx 0.015$ over a wide range of environments (e.g., \citealt{krumholz12, salim15, leroy17}, though see \citealt{murray11, lee16}), which provides a very simple model for star formation and a starting point for our model. 

We will later examine some models for the origin of this relation, but for now we write $\dot{\Sigma}_* $ as a power law function of the disk gas density $\Sigma_g$, normalized by an efficiency parameter $\epsilon_{* ,0}$,
\begin{equation}
\dot{\Sigma}_* = \epsilon_{* ,0} {\Sigma_{g,0} \over t_{\rm ff,0}^{\rm disc}} \left( {\Sigma_g \over \Sigma_{g,0}} \right)^\chi \left( {1+z \over 1 + z_0} \right)^\zeta,
\end{equation}
where we will leave $\chi$ and $\zeta$ as free parameters for now; the explicit redshift dependence allows, for example, for the star formation law to depend on the galaxy's orbital time, etc. (The models we will consider later have $\chi \approx 1$--2.) Our simplest star formation prescription (labeled `KS' for Kennicutt-Schmidt) will take $\chi = 1.4$ and $\zeta=0$, as in equation~(\ref{eq:KSlaw}). 

In this work we wish to integrate over the entire galaxy disc to obtain the total star formation rate. In principle, we can do so by following gas as it flows inward \citep{munoz12,forbes14}. However, following \citet{krumholz18} we will simply average over the disc parameters by multiplying the star formation rate evaluated at the disc's scale radius by a factor $\phi_a \sim 2$, which parameterizes the effects of area-averaging.

%FIGURE: Feedback models
\begin{figure}
	\includegraphics[width=\columnwidth]{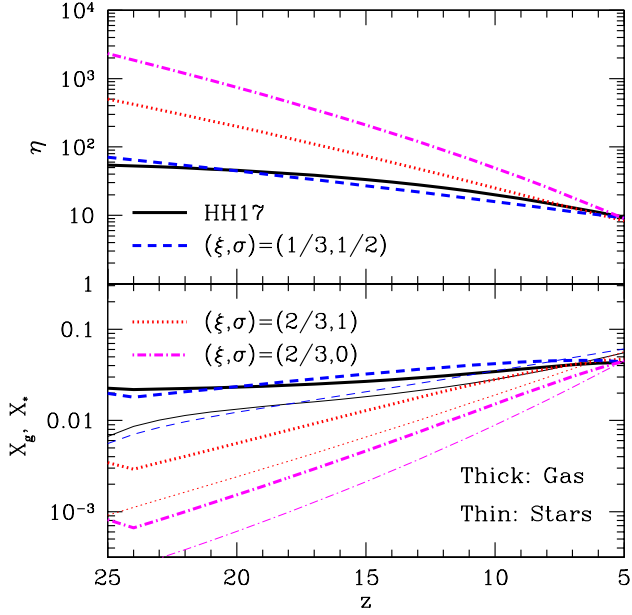} 
    \caption{Bathtub models with varying feedback prescriptions. The short-dashed, dotted, and dot-dashed curves use the power-law prescription of eq.~(\ref{eq:eta-param}) for the feedback law. These correspond to momentum-regulated, energy-regulated, and mass dependence from energy regulation but independent of redshift. The solid  curve uses the inhomogeneous feedback law of \citet{hayward17}, which will be described in section \ref{fig-disc}. In all cases, we have chosen parameters so that they have roughly the same mass-loading parameter at $z=5$.}
    \label{fig:bathtub-feedback}
\end{figure}

\subsection{A simple outflow model} \label{fg-simple}

The simplest approach to model gas outflows is to balance the feedback generated by star formation -- in particular, supernovae -- with a condition for gas to escape the host system. In \citet{furl17-gal}, we balanced the rate at which the accreting gas delivers binding energy with the rate at which supernovae supply feedback energy, which suggests a scaling $\eta \propto m_h^{-2/3} (1+z)^{-1}$. Alternatively, demanding that the momentum provided by supernovae accelerates the remaining gas reservoir to the escape velocity of the halo demands that $\eta \propto m_h^{-1/3} (1+z)^{-1/2}$. For now, we will write a general form that allows for the feedback prescription to vary with halo mass and cosmic time:
\begin{equation}
\eta = \eta_0 \left( {m_h \over m_0} \right)^{-\xi} \left({1+z \over 1 + z_0} \right)^{-\sigma},
\label{eq:eta-param}
\end{equation}
where $\eta_0$, $\xi$, and $\sigma$ are constants. Satisfactory fits to the high-$z$ luminosity functions can be found with reasonable choices for the proportionality constants \citep{furl17-gal}, although the current observations prefer a prescription that is redshift-independent \citep{mirocha17, mirocha20}, indicating that the simplest feedback descriptions are not complete. 

In this framework, the mass-loading parameter evolves as a galaxy accretes more material, through its dependence on halo mass or redshift (or both). Figure~\ref{fig:bathtub-feedback} shows some examples of applying  different feedback prescriptions to the bathtub model. The upper panel shows $\eta$, while the lower panel shows the gas and stellar retention fractions. We have normalized the curves so that they produce roughly the same value of $\eta$ at $z=5$ for ease of comparison. These curves follow a single halo as it grows over time, so the evolution is a combination of both the explicit $z$-dependence and the growing halo mass. Note that, when $\xi \sim 2/3$, the gas retention fraction evolves significantly over this interval. However, the qualitative conclusions  of the bathtub model still apply, because $X_g$ is always very small and the evolution is significantly slower than that of the halo mass accretion rate.

\subsection{Insights from Power-Law Prescriptions} \label{insight-power-law}

With these prescriptions, we can rewrite the basic system of equations for galaxy evolution in terms of the surface densities. In non-dimensional form, and again assuming $\eta \gg 1$, the equations become 
\begin{eqnarray}
\tilde{m}_h' & = & - {\mathcal M}_0 \tilde{m}_h, \\
\tilde{m}_g' & = & \tilde{m}_g {\mathcal M}_0 \left[ -{1 \over X_g} \nonumber \right. \\
& & + \left. \eta_0 \dot{X}_{* ,0} \left( {X_g \over X_{g,0}} \right)^{\alpha_X} \tilde{m}_h^{\alpha_m} \left({1+z \over 1 + z_0} \right)^{\alpha_z}   \right], \label{eq:mgprime} \\
\tilde{m}_* ' & = & {\mathcal M}_0 \dot{X}_{* ,0} \left( {X_g \over X_{g,0}} \right)^{\beta_X} \tilde{m}_h^{\beta_m} \left({1+z \over 1 + z_0} \right)^{\beta_z} \label{eq:mstarprime} \\
X_g' & = & X_g \left( {\tilde{m}_g' \over \tilde{m}_g} - {\tilde{m}_h' \over \tilde{m}_h} \right), \label{eq:Xgprime}
\end{eqnarray}
where we have introduced the dimensionless constants ${\mathcal M}_0 = (\dot{m}_0/m_0) / (H_0 \sqrt{\Omega_m})$ and $\dot{X}_{* ,0} = \dot{m}_{* ,0}/\dot{m}_{g,0}$. Here $\dot{m}_{* ,0} = \dot{\Sigma}_{* ,0} \lambda^2 r_{v,0}^2$ and $X_{g,0} = \Sigma_{g,0} \lambda^2 r_{v,0}^2/\dot{m}_{g,0}$. The exponents are:
\begin{eqnarray}
\alpha_X & = & \chi-1, \\
\alpha_m & = & (\chi - 3\xi - 1)/ 3, \\
\alpha_z & = & 2 \chi - \sigma - 9/2 + \zeta, \\ 
\beta_X & = & \chi, \\
\beta_m & = & (2 + \chi)/3, \\
\beta_z & = & 2 \chi - 9/2 + \zeta.
\end{eqnarray}
%Note that the exponents for $\tilde{m}_* '$ do \emph{not} involve the feedback parameters, though those parameters do affect the gas content. 

These equations have qualitatively similar solutions to those of the bathtub model:

\emph{(i)} First, let us assume that the solutions have $X_g \sim$~constant$\ll 1$. As before, this implies that the star formation rate (plus associated feedback) and baryonic accretion rate roughly cancel. By balancing the two terms in square brackets in equation~(\ref{eq:mgprime}), we see that this requires that $X_g \propto (\dot{\Sigma}_{* ,0} \eta_0)^{-1/\chi}$: just as before, the gas fraction depends on both the normalization of the star formation efficiency (here expressed as $\dot{\Sigma}_{* ,0}$) and the mass-loading parameter. 

\emph{(ii)} Because $X_g$ is nearly constant, the gas and stellar mass components grow nearly exponentially with time.  To see this, consider equation~(\ref{eq:mstarprime}), where the $X_g$ factor is nearly constant while the explicit redshift dependence is mild. Meanwhile, $\beta_m \sim 1$, so the stellar mass grows at a similar rate to the halo mass. We have already argued that the two terms in the square brackets in equation~(\ref{eq:mgprime}) roughly cancel at all redshifts, so we expect $\tilde{m}_g' \propto \tilde{m}_g$. 

\emph{(iii)} This in turn implies that $\mbox{SSFR} \propto (1+z)^{5/2}$ (approximately), nearly independently of the details of the star formation law -- mirroring the growth of the specific mass accretion rate (SMAR) of galaxies, SMAR~$\propto (1+z)^{5/2}$ according to simulation fits. As pointed out by \citet{lilly13} with their gas regulator model (very similar in spirit to the bathtub model) and as part of the semi-empirical model of \citet{behroozi13}, such a relation is quite generic, albeit with an offset between the two. The offset occurs because the star formation rate traces the total gas mass, rather than the cosmological accretion rate, and has an efficiency determined by several parameters with mild redshift dependence. But the offset is always only a factor of a few.

\emph{(iv)} This system, like the bathtub model, also implies $m_* \propto 1/\eta_0$, with no dependence on the normalization of the star formation efficiency. To see this, note that equation~(\ref{eq:mstarprime}) yields $\tilde{m}_* ' \propto \dot{\Sigma}_{* ,0} X_g^{\beta_X}$. But, in the equilibrium state, equation~(\ref{eq:mgprime}) implies $X_g^{\beta_X} \propto X_g^{\alpha_X+1} \propto 1/(\dot{\Sigma}_{* ,0} \eta_0)$, which finally yields $m_* \propto 1/\eta_0$. 

As in the bathtub model, these equations imply that the stellar mass of a galaxy depends only on the feedback model, while the gas content adjusts itself based on the star formation model so as to produce the ``correct" mass in stars. In this picture, measuring the small-scale star formation physics inside galaxies requires measurements of the gas content in addition to the stellar content, because the galaxy-averaged efficiency is driven almost entirely by feedback.  

\section{Gas recycling and chemical evolution} \label{met-rec}

Before considering another layer of complexity for feedback and star formation, we now add two additional physical components to the model.

%%%%%%%%%%%%%%%%%%%FIGURE: Bathtub models: Recycling
\begin{figure}
	\includegraphics[width=\columnwidth]{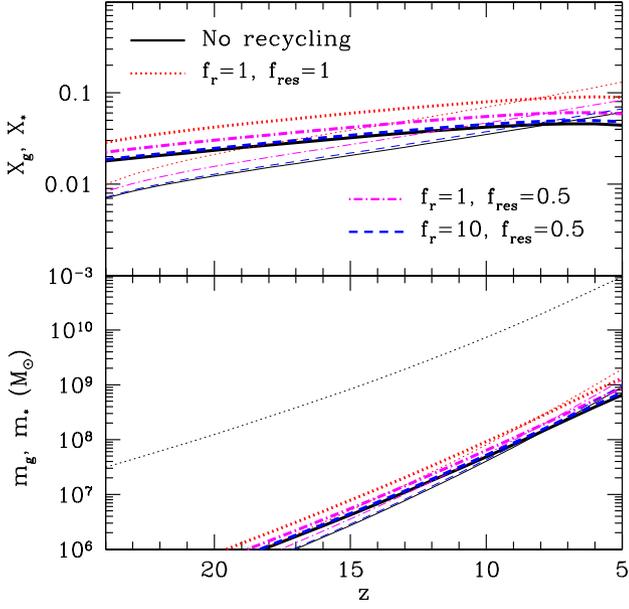} 
    \caption{The effects of gas recycling in bathtub models. The curves follow a halo that ends at $m_h \approx 10^{11} \ M_\odot$ at $z=5$. All assume momentum-regulated feedback.  \emph{Top:} Evolution of the gas retention fraction $X_g$ (thick curves) and the similar star formation efficiency $X_* $ (thin curves).  \emph{Bottom:} The gas and stellar masses (thick and thin curves, respectively), as well as the total halo mass (dotted curve).}
    \label{fig:bathtub-recycle}
\end{figure}

\subsection{The circumgalactic medium and gas recycling} \label{reservoir}

To this point, we have assumed that gas expelled from the galaxy disc leaves the system entirely. However, some of this gas may be ejected at relatively low velocity, and in real galaxies much of it falls back onto the galaxy disc before a long period has elapsed. 

We therefore assume that the ejected gas can follow two paths. A fraction $f_{\rm res}$ will eventually fall back onto the disc. We assume that this gas is re-accreted on a timescale proportional to the dynamical time of the virialized dark matter halo, $t_r = f_r t_{\rm dyn}$. As a fiducial value, we set $f_r=1$. The remainder of the ejected gas escapes the halo entirely, never to be re-accreted. We therefore introduce a new parameter $\eta_{\rm res} \le \eta$ that specifies the rate at which gas is delivered to the reservoir; then gas is ejected entirely from the system with a mass-loading factor $\eta_{\rm out} = \eta - \eta_{\rm res}$. For our power-law outflow model, equation~(\ref{eq:eta-param}), we take $f_{\rm res} = \eta_{\rm res}/\eta$ to be a free parameter. 

We therefore supplement our system of equations with an additional equation tracking the gas reservoir:
\begin{eqnarray}
\dot{m}_g & = & \dot{m}_{c,g} - (\mu+ \eta) \dot{m}_{* } + m_r/t_{r} \label{eq:g-gas-btr} \\
\dot{m}_r & = & \eta_{\rm res} \dot{m}_{* } - m_r / t_{r}. \label{eq:g-recycle-btr}
\end{eqnarray}

Figure~\ref{fig:bathtub-recycle} shows that recycling does not have any signifiant qualitative effects on the models:  it essentially reduces the value of $\eta$, but because the halo is growing rapidly with time anyway, it only changes the stellar and gas masses by factors $\la 2$.

\subsection{Chemical evolution} \label{chem-evol}

These models also allow us to track the chemical evolution of galaxies, so we introduce the metal mass in the galaxy's ISM, $m_Z$. We assume that the stellar population returns a metal mass $y_Z \dot{m}_{* }$ to the ISM over a short timescale, where they must mix with the gas there. In one limit, the metals are mixed uniformly with the gaseous ISM so that the ejected metal mass is simply a fraction $Z$ of the total ejected mass. In the opposite limit, the metals are trapped by the stellar winds and launched immediately out of the galaxy. We capture these limiting cases by assuming that a fraction $f_m$ of the metals produced are injected directly into the wind (c.f.,  \citealt{peeples11, forbes14b}).  Thus, letting $Z = m_Z/m_g$ be the ISM metallicity and letting $m_{r,Z}$ be the mass of metals in the ejected material,
\begin{eqnarray}
\dot{m}_Z & = & -(1+\eta) Z \dot{m}_{* } + (1 - f_m) y_Z \dot{m}_{* } + m_{r,Z}/t_{r}, \label{eq:g-metals} \\
\dot{m}_{r,Z} & = & \eta_{\rm res} Z \dot{m}_{* }  + f_m y_Z (\eta_{\rm res}/\eta) \dot{m}_{* }  - m_{r,Z} / t_{r}, \label{eq:g-recycle-metals} 
\end{eqnarray}
where in the second line we have assumed that the metals do mix uniformly within the hot reservoir and between the gas that is retained by and ejected from the halo. In our calculations, we take $y_Z = 0.03$ \citep{benson10} and assume that solar metallicity corresponds to $Z = 0.0142$ \citep{asplund09}. The effectiveness of direct metal ejection has recently been estimated by \citep{forbes19} using a suite of semi-analytic models. They find that most ($\sim 80\%$) of the metals produced by massive stars are carried away with the wind, albeit with substantial uncertainties. We will consider the cases $f_m=0$ and 0.5 in this work; a larger value will simply decrease the overall metallicity of the galaxies.

\section{Feedback-driven star formation laws} \label{feedback-models}

In section \ref{power-law-model}, we introduced simple power-law descriptions of both star formation and feedback. In this section, we describe a state-of-the-art analytic framework that motivates such relations. In sections \ref{sf-turb} and \ref{sf-rt}, we describe two such models for the star formation law, while in section \ref{fig-disc} we describe a framework for implementing momentum-regulated feedback.

\subsection{Star-forming disc properties} \label{sf-disc}

We first collect some necessary properties of galaxy discs. The fundamental assumptions of many modern star formation models are that: (1) the galaxy disc's vertical support is provided by turbulence stirred -- at least in part -- by stellar feedback and (2) star formation occurs inside molecular clouds, whose own formation is regulated by gravitational instabilities in the disk. The latter can be expressed through the \citet{toomre64} parameter $Q$, which determines whether a region in a rotating disc is locally unstable to gravitational collapse. In detail, the parameter depends on the partitioning of the system into gas, stars, etc. For a gas-dominated disc, the criterion is
\begin{equation}
Q_g = {\kappa \sigma_g \over \pi G \Sigma_g},
\label{eq:qg}
\end{equation}
where $\sigma_g$ is the gas velocity dispersion and $\kappa$ is the epicyclic frequency. For a disk with a flat rotation curve (which we will assume for simplicity), $\kappa = \sqrt{2} \Omega$. 
%defined by $\beta = {\rm d} \ln v_\phi / {\rm d} \ln r$, $\kappa = \sqrt{2 (\beta+1)} \Omega$. 
Real discs are mixtures of stars and gas; the former contribute to the gravitational balance but do not participate in the gravitational collapse. We approximate their effects on stability following \citet{romeo13} and K18 by assuming that they modify the effective $Q$ parameter for marginal stability to $Q = f_{g,Q} Q_g$, where
\begin{equation}
f_{g,Q} = { \Sigma_g \over \Sigma_g + [2 \sigma_g^2/(\sigma_g^2 + \sigma_*^2)] \Sigma_*},
\end{equation}
where $\Sigma_*$ is the surface density of the stellar component and $\sigma_*$ is its velocity dispersion. We assume for simplicity that $\sigma_* = \sigma_g$: although this is certainly too simplistic at $z \la 2$, at high redshifts our discs have so few stars throughout most of their evolution that it makes little difference to the model results. Note that in all of our models the discs are strongly gas-dominated, except possibly at $z \la 7$.

The resulting star formation rates will depend on several properties of the disc. First, the disc free-fall time can be obtained by balancing the vertical components of gravity and pressure, yielding (K18)
\begin{equation}
t_{\rm ff}^{\rm disc} = \sqrt{3 \pi \over 32 G \rho_{\rm g, mp}} = {\pi Q \over 4 f_{g,Q}} \sqrt{ {3 f_{g,P} \phi_{\rm mp} \over 2 } } {1 \over \Omega},
\label{eq:tff-disc}
\end{equation}
where $f_{g,P}$ is a dimensionless factor of order unity parameterizing the contribution of stellar and dark components to vertical equilibrium, and $\phi_{\rm mp}$ is the total midplane pressure relative to the turbulent contribution. For each of these, we take the fiducial values of K18 in our numerical calculations (see their Table~1).\footnote{We note that K18 assume $f_{g,P} = 0.5$, which is not necessarily consistent with our assumptions about the stellar contribution to $f_{g,Q}$: neglecting dark matter, gas-dominated discs would have $f_{g,P} \approx 1$. We set $f_{g,P}=0.5$ as a default, however, because we find dark matter generally provides a larger contribution than in local galaxies. Fortunately, this assumption has only a modest effect, with its largest effect on the gas surface density.} Regardless of these corrections, the free-fall time is comparable to the orbital time.  

For a disc composed of gas and stars, the scale height can be written \citep{forbes12}
\begin{equation}
h \approx {\sigma_g^2 \over \pi G [\Sigma_g + (\sigma_g/\sigma_*) \Sigma_*]} \equiv {\sigma_g^2 \over \pi G \Sigma_g \phi_Q},
\label{eq:scale-height}
\end{equation}
where $\phi_Q$ is another correction factor of order unity.

\subsection{Star formation regulated by stellar feedback} \label{sf-turb}

One conceptually simple limit is if feedback from star formation maintains discs near a point of marginal global stability. As gas accretes onto a disc, it becomes more unstable to gravitational fragmentation and hence star formation. But feedback from the stars injects energy into the medium, increasing the turbulent velocity dispersion $\sigma_g$ until fragmentation slows down. This limiting case, explored in FQH13, results in an especially simple star formation law (see also \citealt{thompson05,munoz12}). We refer the reader to that paper for a more detailed description of the physical inputs and assumptions; see \citet{faucher18} for another simplified treatment and K18 for a discussion of this assumption in the context of more general models. 

FQH13's fundamental ansatz is that the turbulent disc support is provided \emph{entirely} through stellar feedback, which maintains the disc at marginal Toomre stability.  We can therefore derive the star formation rate by balancing the rate at which stellar feedback adds energy to the ISM with that at which turbulence dissipates it. The rate of energy input per unit disk surface area from stellar feedback is
\begin{equation}
\dot{\Pi}_* = \left({P_* \over m_*} \right) \sigma_g \dot{\Sigma}_*,
\label{eq:stellar-input}
\end{equation}
where $P_*/m_*$ is the momentum input rate per unit mass from star formation. Here we have assumed that the supernova shells dissipate into the ISM when they decelerate to a velocity $\sim \sigma_g$, so that they inject energy $\sim P_* \sigma_g$. For the rate of momentum injection, \citet{martizzi15} estimate
\begin{eqnarray}
\left( {P_* \over m_*} \right)_{\rm M15} & = & 3,500 \left( {Z \over 0.1 \ Z_\odot} \right)^{-0.114} \left( {n_{\rm H} \over 1 \ {\rm cm}^{-3}} \right)^{-0.19} \nonumber \\ 
& & \times \left( { \omega_{\rm SN} \over 9.3 \times 10^{48} \ {\rm erg} \, M_\odot^{-1}} \right) \ {\rm km \, s}^{-1} ,
\label{eq:pm-martizzi}
\end{eqnarray}
where $Z$ is the stellar metallicity, $n_H$ is the local ISM density, and $\omega_{\rm SN}$ is the kinetic energy released by supernovae per stellar mass formed.  We have scaled these factors to our fiducial assumptions,\footnote{Given the other uncertainties, we do not vary the assumed momentum injection rate with the galaxy parameters ($n_H$ and $Z$).} but we also introduce an  factor $b_{\rm fb}$ to parameterize the overall strength of feedback relative to equation~(\ref{eq:pm-martizzi}), so that $P_*/m_* \equiv b_{\rm fb} (P_*/m_* )_{\rm M15}$.

For stellar feedback to support the disc through turbulence, the injected energy must balance the rate at which turbulence dissipates it. This energy ($\sim \Sigma_g \sigma_t$ per unit area, where $\sigma_t$ is the turbulent velocity) is lost over roughly the crossing time $t_{\rm eddy}$ of the largest eddy in the turbulent field, which for isotropic turbulence must be of order the disk scale height. This crossing time is therefore $t_{\rm eddy} \sim h/\sigma_t$, and the energy dissipation rate is 
\begin{equation}
\dot{\Pi}_t = \eta {\Sigma_g \sigma_t^2 \over h/\sigma_t} = {2 \over \pi G}  f_T \phi_Q \phi_{\rm nt}^{3/2} f_{g,Q}^2 {\Omega^2 \sigma_g^3 \over Q^2},
\label{eq:turb-diss}
\end{equation}
where $f_T \sim 1$ is an unknown factor and $\phi_{\rm nt} = \sigma_t/\sigma_g$. 

Balancing these two rates, and using equation~(\ref{eq:sfr-breakdown}), we obtain an expression for the required pressure support, written in terms of the gas velocity dispersion:
\begin{equation}
\sigma_g = {4 \over \pi \sqrt{3}} (\eta \phi_Q \phi_{\rm nt}^{3/2} \phi_{\rm mp}^{1/2} f_{g,P}^{1/2} )^{-1} \epsilon_{\rm ff}^{\rm gal} \left({P_* \over m_*} \right).
\label{eq:sigma-sf}
\end{equation}
In combination with equation~(\ref{eq:qg}), we see that the large range in $\Sigma_g$ amongst observed galaxies must translate into either substantial variation in $Q$  or in $\sigma_g$. The latter implies that at least one of the factors on the right-hand side of equation~(\ref{eq:sigma-sf}) must vary -- in particular, unless the assorted non-dimensional factors all change dramatically with galaxy properties, the star formation efficiency $\epsilon_{\rm ff}^{\rm gal}$ must vary significantly. 

Thus, if galaxy discs remain in a condition of marginal gravitational stability, the required efficiency is (see eq.~54 in K18):
\begin{equation}
\epsilon_{\rm ff}^{\rm gal} = {\pi \sqrt{3} \over 4} (\eta \phi_Q \phi_{\rm nt}^{3/2} f_{g,P}^{1/2} \phi_{\rm mp}^{1/2} ) \left({P_* \over m_*} \right)^{-1} \sigma_g.
\label{eq:eps-sf-Q=1}
\end{equation}
Inserting this expression into equation~(\ref{eq:sfr-breakdown}), we find
\begin{equation}
\dot{\Sigma}_* = \pi G \eta f_{g,Q} \phi_Q \phi_{\rm nt}^{3/2} \left( {P_* \over m_*} \right)^{-1} \Sigma_g^2.
\label{eq:fg-sfr}
\end{equation}
The same scaling holds even if one assumes a turbulent ISM in which $Q$ varies across the disc, so that only some regions are unstable (FQH13). We also note that assuming that $\epsilon^{\rm gal}_{\rm ff}$ is constant, but allowing $Q$ to vary, leads to the same scaling in the star formation law \citep{ostriker11}. The key assumption in all cases is that star formation is the \emph{only} mechanism generating the turbulence that supports the disc.

\subsection{Star formation regulated by radial transport} \label{sf-rt}

The FQH13 approach, in which stellar feedback provides \emph{all} the turbulent support for the gas disc, remains controversial. K18 argue that the steepness of the SFR law in these models compared to the observed relation (cf. eq.~\ref{eq:fg-sfr} and \ref{eq:KSlaw}), together with their requirement that the star formation efficiency increase in high-density regions, make such an approach untenable. According to the K18 models, stellar feedback plays a role in supporting discs but is insufficient for most star-forming galaxies because the star formation efficiency per free-fall time within molecular clouds, $\epsilon_{\rm int}^{\rm GMC}$, is fixed at $0.015$, placing an upper limit on the amount of support that supernovae can provide. In this picture, the remainder of the turbulent support is generated by non-axisymmetric instabilities triggered by radial transport of gas through the galaxy disc \citep{krumholz10}. 

In this case, the maximum surface density that can be supported by star formation is
\begin{equation}
\Sigma_{g,sf}^{\rm max} = {8 \sqrt{2} \over \sqrt{3} \pi G} {\epsilon_{\rm ff}^{\rm gal} f_{g,Q} \over Q \eta \phi_{\rm mp} \phi_Q \phi_{\rm nt}^{3/2} f_{g,P}^{1/2} } \left( {P_* \over m_* } \right)  t_{\rm orb}^{-1}.
\label{eq:Sigma-max}
\end{equation}
Whenever the surface density exceeds this value, the remaining support is provided by gravitational inflow, so that the star formation rate increases less rapidly with surface density than otherwise expected.

In this picture, K18 write the surface density of the star formation rate as
\begin{equation}
\dot{\Sigma}_* = f_{\rm sf} {\Sigma_g \over t_{\rm sf,rt}},
\label{eq:sigstar-k18}
\end{equation}
where $f_{\rm sf}$ is the fraction of the gas able to form stars and the star formation timescale is $t_{\rm sf,rt} = t_{\rm ff}^{\rm disc}/ \epsilon_{\rm int}^{\rm GMC}$, where $\epsilon_{\rm int}^{\rm GMC}=0.015$ by fiat.\footnote{K18 also demand that $t_{\rm sf,rt} < t_{\rm sf,max} = 2$~Gyr, the value found in galaxies like the Milky Way where the molecular gas breaks into discrete clouds. High-$z$ galaxies have sufficiently high surface densities that this restriction will not be important.} For now, we can take $f_{\rm sf}=1$, but we will revisit this parameter in the next section. 

\subsection{Outflows in a turbulent gas disc} \label{fig-disc}

The same picture of turbulence stirred by stellar feedback (or other sources) also provides an elegant method to estimate the effects of outflows \citep{thompson16,hayward17}. In particular, we imagine a turbulent, inhomogeneous ISM and assume that gas can be blown out of the disc if stellar feedback is sufficiently strong on a local level. 

In particular, if supernova blast waves cool and lose their thermal energy in the disc, \citet{hayward17} argue that the relevant criterion for gas to escape is whether the momentum carried by radiation and supernova blastwaves can accelerate material to the disc escape velocity within one disc turbulent coherence time $t_{\rm eddy} \sim t_{\rm orb} \sim r_d/v_c$ (after which the density field is assumed to reset, scrambling the momentum injection). In the model, the variable column density induced by turbulence means that the feedback accelerates each patch to a different terminal velocity. Only regions with a terminal velocity exceeding the disc escape velocity are lost. In this picture, the condition for a gas parcel to escape the galaxy is\footnote{As in \citet{hayward17}, we implicitly assume that the momentum injected from stellar feedback is uniformly spread throughout the galaxy disc, even though most star formation is occurring in areas with large $\Sigma_g$.}
\begin{equation}
\dot{\Sigma}_* \left( {P_* \over m_*} \right) > \Sigma_g {v_{\rm esc} \over t_{\rm eddy}} \rightarrow 
\Sigma_g < \Sigma_g^{\rm max,p} = \left( {P_* \over m_*} \right) {\dot{\Sigma}_* \over \sqrt{2} v_c \Omega_d},
\label{eq:turb-cond}
\end{equation}
where $v_{\rm esc} = \sqrt{2} v_c$ is the local escape velocity. 

 We therefore require a model for the turbulent density field, which we take from \citet{hayward17}. Here, the probability that a mass element is in a region with surface density $\Sigma$ is lognormal,
\begin{equation}
p_{\rm ISM}(x) = {1 \over \sqrt{2 \pi \sigma_{\ln \Sigma}^2}} \exp \left( - { (x - \bar{x})^2 \over 2 \sigma^2_{\ln \Sigma}} \right),
\label{eq:pdist-turb}
\end{equation}
where $x = \ln (\Sigma / \VEV{\Sigma} )$, $\bar{x} = \sigma^2_{\ln \Sigma}$,  $\sigma^2_{\ln \Sigma} \approx \ln (1 + R {\mathcal M}^2/4)$, where the Mach number is ${\mathcal M} = \sigma_g/c_s$ and $R$ is a function of the Mach number and the shape of the turbulent power spectrum; see eq. (14) of \citet{thompson16} for details. To estimate the Mach number, we assume that the sound speed is $c_{s,4} \approx 12$~km/s, appropriate for primordial gas at $T \approx 10^4$~K. However, in some very small haloes we can have $\sigma_t < c_{s,4}$. We therefore assume that photoheating from H~II regions can also drive turbulence and take ${\mathcal M} = (c_{s,4}^2 + \sigma_t^2)^{1/2}/c_{s,4}$. (Our results are not sensitive to the details of these choices.)

The fraction of the ISM that is ejected per coherence time is then
\begin{equation}
f_{\rm out} = \int_{-\infty}^{x_{\rm out}} p_{\rm ISM}(x) dx = {1 \over 2} \left[ 1 - {\rm erf} \left({-2 x_{\rm out} + \sigma^2_{\ln \Sigma} \over 2 \sqrt{2} \sigma_{\ln \Sigma}} \right) \right], 
\label{eq:fout}
\end{equation}
where $x_{\rm out} = \ln (\Sigma_{g,{\rm max}} / \VEV{\Sigma_g} )$.

Finally, the mass-loading factor is 
\begin{equation}
\eta = { f_{\rm out} \VEV{\Sigma_g} \Omega_d \over \dot{\Sigma}_*}.
\end{equation}
To estimate the mass fraction that is ejected entirely from the system, we assume that gas able to escape the halo entirely must be provided a momentum per unit mass of $f_{\rm ej} v_{\rm esc}$, where $f_{\rm ej}>1$. As a fiducial value, we set $f_{\rm ej} = 3$.  In this case, $\eta_{\rm res}$ is composed of gas imparted enough energy to escape the disc but not enough to meet this more stringent threshold.

Although the physics of the turbulent outflow model is considerably deeper than the parameterized form of equation~(\ref{eq:eta-param}), it nevertheless approximately reduces to that form if $\eta \gg 1$. To see this, let us consider the star formation prescriptions in sections \ref{sf-turb} and \ref{sf-rt}. In the former case, $\dot{\Sigma}_* \propto \Sigma_g^2$, so $\Sigma_g^{\rm max}/\VEV{\Sigma_g} \propto (\dot{\Sigma}_* t_{\rm orb}) / (\Sigma_g v_c) \propto m_h/(r_d v_c^2)$. But $r_d \propto m_h^{1/3} (1+z)^{-1}$ and $v_c \propto m_h^{1/2}/r_d$, so this is roughly constant -- in reality, $x_{\rm out}$ and $f_{\rm out}$ vary slowly in such models, with $f_{\rm out} \sim 1/2$. Then, $\eta \propto f_{\rm out} \Sigma_g \Omega_d/\dot{\Sigma}_* \propto m_h^{-1/3} (1+z)^{-1/2}$. Thus the mass and redshift scaling are identical to the case of naive momentum-regulation, at least in the limit of strong outflows and discs supported purely through stellar feedback. Figure~\ref{fig:bathtub-feedback} illustrates how similar this model is to the naive approach.

In the model with radial transport, $\dot{\Sigma}_* \propto \Sigma_g/t_{\rm orb}$ (assuming for simplicity that $f_{\rm sf} \rightarrow 1$). Then $\Sigma_g^{\rm max}/\VEV{\Sigma_g} \propto m_h^{-1/3} (1+z)^{-1/2}$, which means that this parameter decreases rapidly as halos grow (although the actual fraction of ejected gas varies more slowly, because $x_{\rm out}$ depends on the logarithm of the ratio). However, in the mass-loading factor, the ratio $\Sigma_g \Omega_d/\dot{\Sigma}_*$ is nearly constant in this model. Thus $\eta \propto f_{\rm out}$, which carries all the mass and redshift dependence in this regime.

%However, in the small galaxies we will study, it is not obvious that all supernovae will lose their energy before blowing out of the galaxy disc. To very approximately model this possibility in small, high-$z$ galaxies, we also consider a model in which we follow a similar argument to \citet{hayward17}, except that the criterion for mass loss is that the energy input into a column of gas over $t_{\rm eddy}$ be sufficient to unbind the gas. If, as before, the feedback occurs uniformly through an inhomogeneous medium, this occurs if 
%\begin{equation}
%\dot{\Sigma}_* \omega_{\rm SN} > \Sigma_g {v_{\rm esc}^2/2 \over t_{\rm eddy}} \rightarrow 
%\Sigma_g^{\rm max,E} = {\dot{\Sigma}_* \omega_{\rm SN} \over  v_c^2 \Omega_d},
%\label{eq:turb-cond-energy}
%\end{equation}

\section{The star-forming gas phase} \label{sf-phase}

The final layer of complexity in our star formation model is to distinguish between two ISM phases, one of which forms stars and one of which acts as a warmer gas reservoir. Note that this is not a necessary component: FQH13, for example, do not include it, assuming that the marginal stability condition will essentially force the star-forming phase to grow until enough stars can be formed. 

\subsection{Star formation in molecular gas} \label{fmol}

One common criterion for star-forming gas is that it be molecular. We implement this approximately by following \citet{krumholz13}, who consider the formation of molecular hydrogen in galaxies (building upon \citealt{krumholz09a,krumholz09b}). The molecular fraction is governed by the balance between the interstellar radiation field (which suppresses molecule formation) and self-shielding of the clouds (which depends on the cloud column density and metallicity). 

The model has two simple limits, which describe the full \citet{krumholz13} results accurately enough for our purposes.\footnote{The approximation can be improved in the transition regime between these two limits; we do not attempt to do here largely to illustrate that even the more abrupt threshold condition in the simplified model still has no obvious effect on the overall galaxy evolution.)} Assuming chemical equilibrium between destruction of H$_2$ by Lyman-Werner photons and creation on dust grains, the molecular fraction is governed by the dimensionless parameter
\begin{equation}
\chi = {f_{\rm diss} \sigma_{\rm d} c E_0^* \over n_{\rm CNM} {\mathcal R}},
\end{equation}
where $f_{\rm diss} \approx 0.1$ is the fraction of absorptions of the Lyman-Werner photons by H$_2$ molecules that result in dissociation, $\sigma_d \approx 10^{-21} (Z/Z_\odot)$~cm$^{-2}$ is the dust absorption cross section per hydrogen nucleus for Lyman-Werner photons, $E_0^* = 7.5 \times 10^{-4} G'_0$~cm$^{-3}$ is the number density of Lyman-Werner photons, $G'_0$ is the intensity of the interstellar radiation field in units of the local Solar value, $n_{\rm CNM}$ is the number density of the cold neutral phase, and ${\mathcal R} \approx 10^{-16.5} (Z/Z_\odot)$~cm$^3$~s$^{-3}$ is the rate coefficient for H$_2$ formation on the surface of dust grains. The H$_2$ fraction -- which in this model we associate with the fraction of the ISM in the star-forming phase, $f_{\rm sf}$ -- can then be approximated by \citep{mckee10}
\begin{equation}
f_{\rm sf} = 1 - (3/4) s / (1 + s/4) 
\end{equation}
for $s < 2$ or zero otherwise. Here 
\begin{equation}
s \approx {\ln ( 1 + 0.6 \chi + 0.01 \chi^2) \over 0.6 \tau_c},
\label{eq:sdefn}
\end{equation}
and
\begin{equation} 
\tau_c = 0.066 f_c \left( {\Sigma_g \over 1 \ M_\odot \ {\rm pc}^{-2}} \right) \left( Z \over Z_\odot \right).
\end{equation}
Finally, $f_c$ is the ratio between the surface densities of the cold phase and the surface density averaged over our galactic disks. Following \citet{krumholz13}, we take $f_c=5$, although there are no observations of this clumping factor in high-$z$ galaxies. We also note that the model will break down at very low metallicities because the H$_2$ cannot then form quickly enough; \citet{krumholz13} estimate that this occurs at $Z \la 0.01 Z_\odot$, which would affect the earliest phases of star formation inside galaxies. 

In the high surface density regime \citep{wolfire03},
\begin{equation}
\chi \approx 3.1 \left( {1 + 3.1 (Z/Z_\odot)^{0.365} \over 4.1} \right).
\end{equation}
In practice, this leads to a molecular fraction that is zero below a (metallicity-dependent) threshold (corresponding to $s=2$ in equation~\ref{eq:sdefn}) and then rises rapidly. 

However, below this threshold we must account for the disc pressure support. In this limit, for which $f_{\rm sf} \ll 1$, \citet{krumholz13} shows that the fraction approaches
\begin{equation}
f_{\rm sf} \approx {1 \over 3} \left( 2 - {440 \ {\rm Gyr} \over f_c (Z/Z_\odot) n_{\rm CNM,hs}} \, {\dot{\Sigma}_* \over \Sigma_g} \right),
\end{equation}
where $n_{\rm CNM,hs}$ is the density of the cold medium in this limiting case: it is determined by the balance between the vertical component of gravity and the thermal pressure of the cold medium. In the limit in which the dark matter and/or stars dominate the local density, $n_{\rm CNM,hs} \propto \Sigma_g$. Then, if we write $\dot{\Sigma}_* \propto f_{\rm sf} \Sigma_g$, as in equation~(\ref{eq:sigstar-k18}) and rearrange, we find that $f_{\rm sf} \propto \Sigma_g$ in the limit in which the fraction is small  (see Fig.~1 of \citealt{krumholz13}). Thus, from equation~(\ref{eq:sfr-breakdown}), we see that at low surface densities $\dot{\Sigma}_* \propto \Sigma_g^2$, while above the transition threshold (where $f_{\rm sf}$ saturates), $\dot{\Sigma}_* \propto \Sigma_g$. 

%%%%%%%%%%%%%%%%%FIGURE: SFR Laws
\begin{figure}
	\includegraphics[width=\columnwidth]{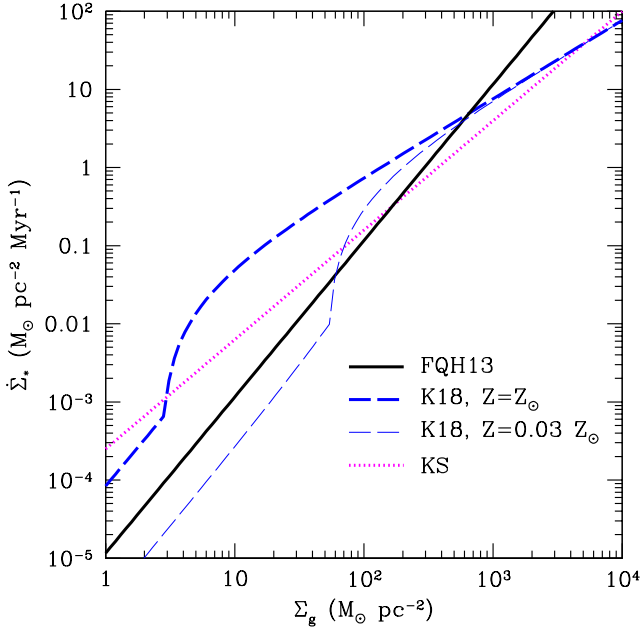} 
    \caption{Comparison of star formation laws for our different prescriptions. The solid black curve is the empirical Kennicutt-Schmidt law. The long-dashed curves apply the model to a $10^{11} \ M_\odot$ halo at $z=8$, taking $Z = (1,\,0.03) \ Z_\odot$, respectively. Note that the curves are normalized to be similar in the surface density range most relevant to our models $\Sigma_g \sim 10^2$--$10^3 \ M_\odot$~pc$^{-2}$.}
    \label{fig:sfr-laws}
\end{figure}

Figure~\ref{fig:sfr-laws} compares the star formation laws resulting from these different $\Sigma_g$-based prescriptions. Note that the K18 model depends on the galaxy properties beyond the surface density; here we apply it to a $10^{11} \ M_\odot$ halo at $z=8$, taking $Z = (1,\,0.03) \ Z_\odot$ for the thick and thin curves. In the K18 models, the sharp feature is induced by the transition to efficient H$_2$ formation. Because this process is driven by dust grains, the surface density at which it occurs depends on metallicity: $f_{\rm sf}$ increases only at significantly larger column densities in the thin curve (though note that at such low metallicities H$_2$ formation is so slow that the model may be breaking down; \citealt{krumholz13}). 

We must emphasize that the normalization of these star formation laws is quite uncertain, as both the FQH13 and K18 scenarios have several unknown parameters. Here we have used the fiducial choices of K18 for simplicity. Below, we will find that high-$z$ galaxies are always in the high-surface density limit, $\Sigma_g \ga 100 \ M_\odot$~pc$^{-2}$.

\subsection{Star formation regulated by gas cycling}

An alternative approach to defining the star-forming phase is based on \citet{semenov17, semenov18}, who appeal to processes that cycle gas between a diffuse phase and the star-forming gas. The model is agnostic to the properties of that star-forming phase (other than requiring that it is self-gravitating and thus subject to collapse), and it does not explicitly require the star-forming gas to be molecular. We write the mass fluxes $\dot{m}_{g \rightarrow {\rm sf}}$ and $\dot{m}_{{\rm sf} \rightarrow g}$ to describe the flow between the star-forming gas (whose mass is $m_{\rm sf}$) and the diffuse ISM (for which we use $m_g$). In this picture, local feedback from young stars as well as dynamical processes rapidly dissipate clouds, before much of their gas is able to form stars. But dynamical and thermal instabilities regenerate star-forming clouds, and it is the balance between these processes that regulates star formation. We obtain this system of equations for the phases of baryonic matter (diffuse gas, star-forming gas, ejected gas, and stars) in and around the galaxy:
\begin{eqnarray}
\dot{m}_g & = &  - \dot{m}_{g \rightarrow {\rm sf}} + \dot{m}_{{\rm sf} \rightarrow g} - \eta \dot{m}_{*} + m_r/t_{r} \label{eq:g-gas} \\
\dot{m}_{\rm sf} & = & \dot{m}_{g \rightarrow {\rm sf}} - \dot{m}_{{\rm sf} \rightarrow g} - \mu \dot{m}_{*} \label{eq:sem-fstar} \\
\dot{m}_{*,{\rm tot}} & = & \mu \dot{m}_{*} \label{eq:g-star} \\
\dot{m}_r & = & \eta_{\rm res} \dot{m}_{*,{\rm ISM}} - m_r / t_{r}. \label{eq:g-recycle}
\end{eqnarray}

The keys to this picture are the timescales governing how gas moves between the various phases. As before, we let $f_{\rm sf} = m_{\rm sf}/m_g$ be the fraction of the galaxy's gas mass in the star-forming phase. S18 parameterize the mass fluxes as 
\begin{eqnarray}
\dot{m}_{g \rightarrow {\rm sf}} & = & { (1-f_{\rm sf})  m_g \over \tau_{g \rightarrow {\rm sf}} }, \\
\dot{m}_{{\rm sf} \rightarrow g} & = & \dot{m}_{{\rm sf} \rightarrow g}^{\rm fb} + \dot{m}_{{\rm sf} \rightarrow g}^{\rm dyn}, \\
\dot{m}_{{\rm sf} \rightarrow g}^{\rm fb} & = & \xi \dot{m}_*, \\
\dot{m}_{{\rm sf} \rightarrow g}^{\rm dyn} & = & { f_{\rm sf}  m_g \over \tau_{{\rm sf} \rightarrow g}^{\rm dyn}}, \\
\dot{m}_* & = & { m_{\rm sf} \over \tau_{*}},
\end{eqnarray}
where $\dot{m}_{{\rm sf} \rightarrow g}^{\rm fb}$ is the rate at which star-forming clouds are destroyed by stellar feedback, $\dot{m}_{{\rm sf} \rightarrow g}^{\rm dyn}$ is the rate at which dynamical processes shred clouds, $\xi$ is the mass-loading factor from feedback (defined \emph{locally} within clouds, rather than integrated over an entire galaxy), and we have introduced three timescales ($\tau_{g \rightarrow {\rm sf}}$, $\tau_{{\rm sf} \rightarrow g}^{\rm dyn}$, and $\tau_*$) to parameterize how fast the processes occur. Note that we have now defined the star formation rate through the mass in this gas phase, so we have introduced a new timescale $\tau_*$ to describe how rapidly this ``star-forming" gas is turned into stars.

We take $\tau_{*} = \tau_{\rm ff}^{\rm disc}/\epsilon_{\rm int}^{\rm GMC}$ for this timescale.\footnote{Note the analogy here to the bathtub model in the star-formation timescale definition -- but the S18 model applies that timescale to individual clouds rather than the galaxy as a whole, and includes additional physics regulating each cloud.} Here, $\tau_{\rm ff}^{\rm disc}$ is the free-fall time (which we assume to apply to molecular clouds as well)\footnote{S18 show that, in Milky-Way like systems, the free-fall time is a few million years and not particularly sensitive to the star formation and feedback prescriptions. However, at very early times this can still be longer than the dynamical time of galaxy-sized halos. We therefore use equation~(\ref{eq:tff-disc}) for this timescale.} and $\epsilon_{\rm int}^{\rm GMC}$ is the star formation efficiency (i.e., the fraction of cloud mass turned into stars per free-fall time); as in the K18 model, we take $\epsilon_{\rm int}^{\rm GMC}=0.015$ as a fiducial value.

Star-forming clouds form through turbulence, cooling, and large-scale gravitational instabilities, a combination of which fix $\tau_{g \rightarrow {\rm sf}}$. S18 find that the resulting timescale is comparable to the orbital time of their simulated galaxy (see also \citealt{semenov17}). We therefore write the relevant cloud formation timescale as $\tau_{g \rightarrow {\rm sf}} =  C_{g \rightarrow {\rm sf}} t_{\rm ff}^{\rm disc}$, where $C_{g \rightarrow {\rm sf}}$ is an unknown parameter. We take $C_{g \rightarrow {\rm sf}} = 1$ as a fiducial value.

Feedback drives gas from the star-forming clouds back to the diffuse ISM through photoionization, winds, and especially supernovae. In their simulations, S18 calibrated their local feedback input to \citet{martizzi15}, corrected for numerical losses, and found $\xi \approx 60$ for their fiducial parameters. We scale this parameter in the same way they did.  Note that, although this parameter is related to the mass-loading parameter $\eta$, it describes feedback on a different physical scale -- how gas is ejected from star-forming clouds into diffuse ISM, rather than out of the galaxy.

The timescale $\tau_{{\rm sf} \rightarrow g}^{\rm dyn}$ represents the rate at which dynamical processes like turbulent shear and differential rotation disperse clouds. S18 find that $\tau_{g \rightarrow {\rm sf}}/\tau_{{\rm sf} \rightarrow g}^{\rm dyn} \sim 4$ in their simulations, i.e. clouds are disrupted quite rapidly relative to their formation. However, this calibration is measured in galaxies similar to the Milky Way, where star formation is relatively slow and the star-forming phase is subdominant. Our galaxies are in a much different regime, and furthermore their dynamical times are far smaller than that of the Milky Way. We therefore set $\tau_{g \rightarrow {\rm sf}}/\tau_{{\rm sf} \rightarrow g}^{\rm dyn} = C_{\rm dyn}$, assuming that these processes remain linked but varying the relative timescale. As a fiducial value, we take $C_{\rm dyn} = 1/4$, which we find matches more closely to the other star formation models we consider in the relevant regime, as we will show in the next section. In practice, this means that dynamical processes form star-forming clouds much more rapidly than they disperse them, which may reflect the higher surface densities of our discs compared to Milky Way analogs.

In this model, galaxies typically approach a ``quasi-equilibrium" approximation in which the mass fraction in the star-forming phase changes only slowly (relative to the dynamical timescales), so that by equation~(\ref{eq:sem-fstar}) we have
\begin{equation}
\dot{m}_* \approx \dot{m}_{g \rightarrow {\rm sf}} - \dot{m}_{{\rm sf} \rightarrow g}.
\end{equation}
This is analogous to the slow growth of the ISM in the bathtub model across cosmological timescales. In this approximation, the galaxy-wide gas depletion time $\tau_{\rm dep} = m_g/\dot{m}_*$ can be written (S18)
\begin{equation}
\tau_{\rm dep} = (1+\xi) \tau_{g \rightarrow {\rm sf}} + \left( 1 + C_{\rm dyn} \right) {\tau^{\rm disc}_{\rm ff} \over \epsilon_{\rm GMC}^{\rm int}}.
\label{eq:taudep}
\end{equation}
This expression offers powerful insights into the processes regulating star formation in galaxies. The first term represents a picture in which clouds form on a timescale $\tau_{g \rightarrow {\rm sf}}$ and are fully ``processed" into stars, subject only to their feedback. In that case, only a fraction $1/(1+\xi)$ of the gas is actually turned into stars, with the rest ejected into the surrounding medium (from which it must again fragment into clouds to form more stars). Any given gas parcel must therefore cycle into clouds $(1+\xi)$ times before it is transformed into stars. 

But this is not the full picture, because clouds do not immediately process their stars. Instead, only a fraction $\epsilon_{\rm GMC}^{\rm int}$ is transformed per free-fall time. Thus, even in the absence of feedback, the expected depletion time of any given cloud is $\tau^{\rm disc}_{\rm ff} /\epsilon_{\rm GMC}^{\rm int}$, which is increased by a factor $(1 + C_{\rm dyn})$ because of dynamical destruction.

While starburst galaxies are typically assumed to be in the ``feedback-limited" regime, equation~(\ref{eq:taudep}) reveals an important wrinkle: if dynamical processes disrupt clouds on times comparable to the orbital time, the second term cannot be ignored. At high redshifts, clouds are processing gas into stars on similar timescales to the rate at which the clouds themselves are forming. Given the assumed inefficiency of star formation on that timescale, both the feedback parameter $\xi$ and the cloud-scale star formation efficiency $\epsilon_{\rm GMC}^{\rm int}$ are important to this model -- this is in contrast to the FQH13 model, in which the assumption that feedback supports the disc eliminates the dependence on the efficiency parameter. In the context of the gas cycling picture, the FQH13 assumption essentially requires that  either clouds themselves form more efficiently or that they form stars more efficiently than their local analogs. This is also why we choose a small fiducial value for $C_{\rm dyn}$: otherwise dynamical destruction of clouds essentially caps the star formation efficiency.

In either case, in the limit in which equation~(\ref{eq:taudep}) applies, the gas cycling picture reduces approximately to the bathtub model. It implies $\dot{m}_* \approx m_g/[ K\tau_{g \rightarrow {\rm sf}}]$, where $K$ is a constant in our model. This has the same form as in the bathtub equations.

We emphasize that the S18 model takes a very different perspective on star formation than the other models we have described, as it does not attempt to connect star formation to any of the global galaxy parameters. As a result, for example, the star formation rate is only indirectly a function of the gas surface density. It should be possible to connect the S18 picture more closely with ``global" models, but we do not do so here because considering a range of contrasting models of star formation helps us to understand its robust and generic properties.

\section{Evolution of a disc galaxy} \label{gal-zevol}

Having now reviewed a variety of approaches to describing star formation in galaxies, we next examine how individual galaxies evolve in these models; in section \ref{mass-trends}, we will consider how the results depend on the galaxy's mass and redshift.

%%%%%%%%%%%%%%%%%FIGURE: fstar calibration
\begin{figure}
	\includegraphics[width=\columnwidth]{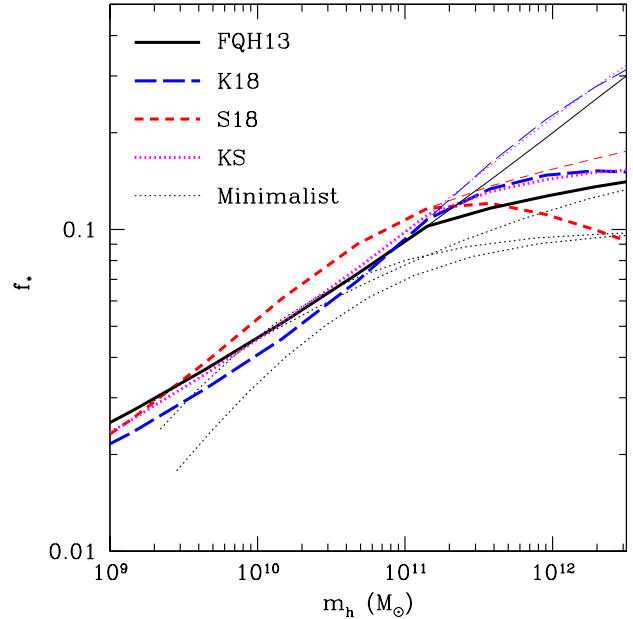} 
    \caption{The halo mass-star formation efficiency relation in our models, where $f_* = \dot{m}_*/\dot{m}_{c,g}$, at $z=7$. The darker thin dotted lines show the models of \citet{furl17-gal}, which provide a reasonable match to observed luminosity functions. The thin lines are the same relation in our models, where we have adjusted the feedback parameter in each case to provide a reasonable match at low masses (see text). The thick curves include a correction reducing the accretion rate onto very massive haloes.}
    \label{fig:fstar}
\end{figure}

\subsection{Choosing our model parameters} \label{lf-fit}

While we will not rigorously test these star formation models against observations, we do want to ensure that they are reasonable analogs to real systems. All of our fiducial models will use the \citet{hayward17} feedback prescription, but we include several star formation prescriptions: (1) the FQH13 star formation law, in which stellar feedback supports the disc (shown by solid lines in most figures); (2) the K18 model, in which radial transport supports the disc at high surface densities and star formation only occurs in molecular gas (long-dashed lines); (3) the gas cycling model of S18 (short-dashed lines); and (4) a simple model in which the star formation is assumed to follow the \citet{kennicutt98} relation (dotted lines). 

The most incisive tests of galaxy formation at high redshifts remain luminosity functions, of which there are now many measurements with reasonably good statistics at the bright end  (e.g., \citealt{bradley12, mclure13, bouwens15, finkelstein15, mcleod16, morishita18, bouwens19, bowler17, bowler20}). Because the UV luminosity traces recent star formation, these measurements essentially provide constraints on the star formation rate as a function of halo mass.

In \citet{furl17-gal}, we compared a ``minimalist" model of galaxy formation centered around the feedback models of section \ref{fg-simple} to these luminosity functions. We found that such a model provided reasonably good matches to the observations. The thin dotted curves in Figure~\ref{fig:fstar} shows the instantaneous star formation efficiency, $f_* = \dot{m}_*/\dot{m}_{c,g}$, in three such models: one assumes momentum-regulated feedback, while the others (with steeper mass dependence at $m_h \la 10^{11} \ M_\odot$) assume energy regulation. 

%%%%%%%%%%%%%%%%%FIGURE: Basic galaxy model: star formation laws
\begin{figure*}
	\includegraphics[width=\columnwidth]{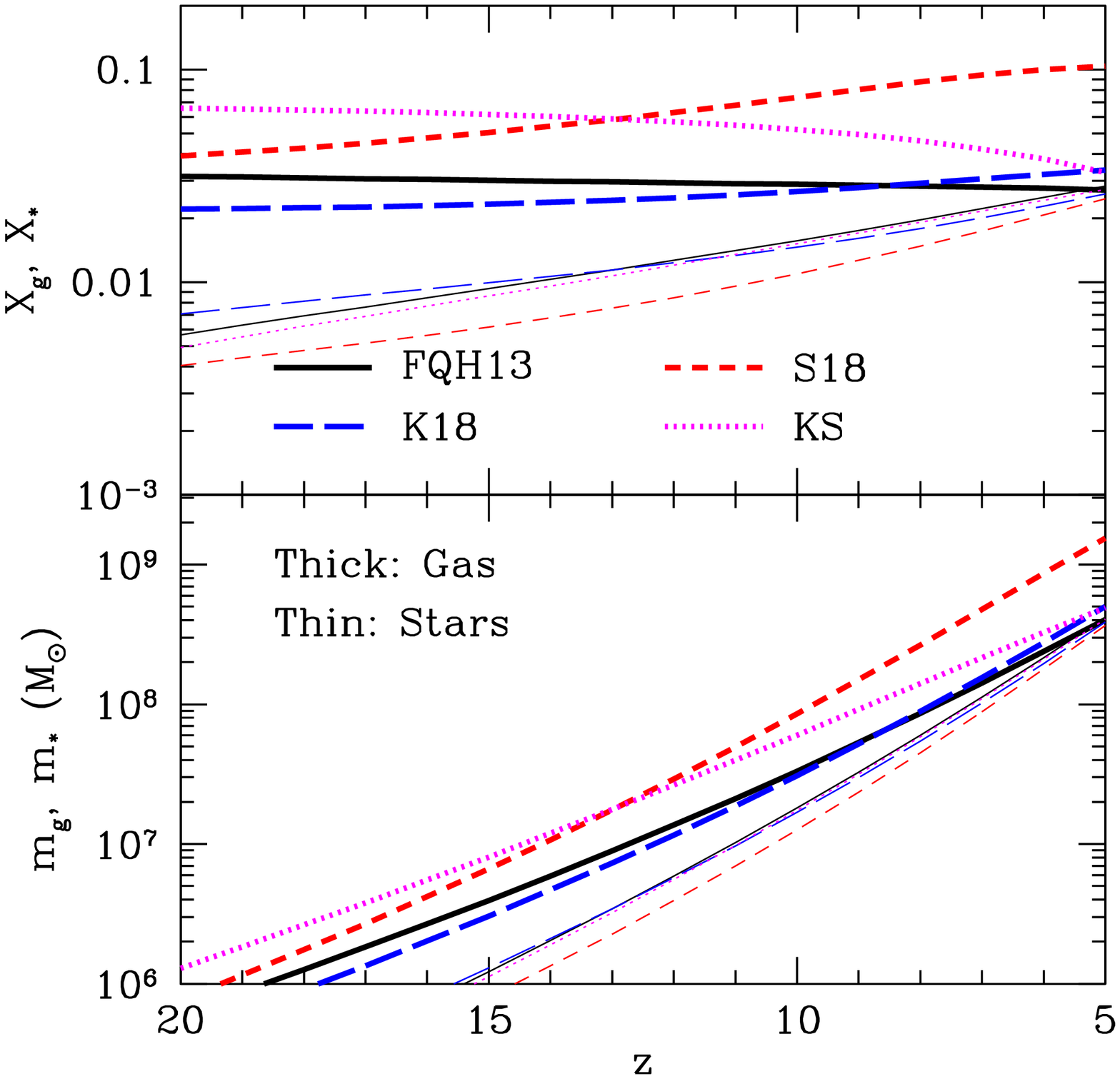} 
	\includegraphics[width=\columnwidth]{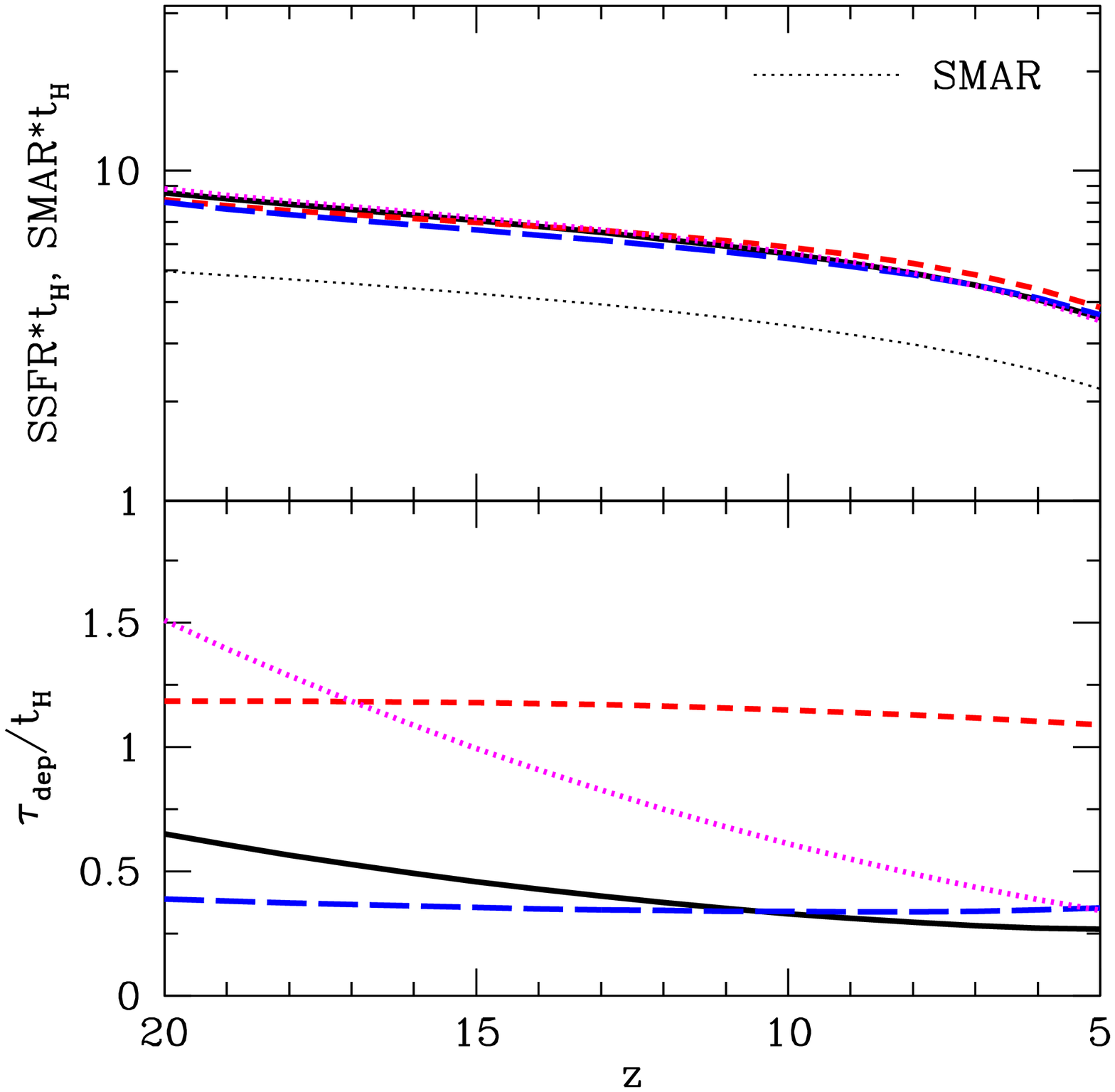} \\
	\includegraphics[width=\columnwidth]{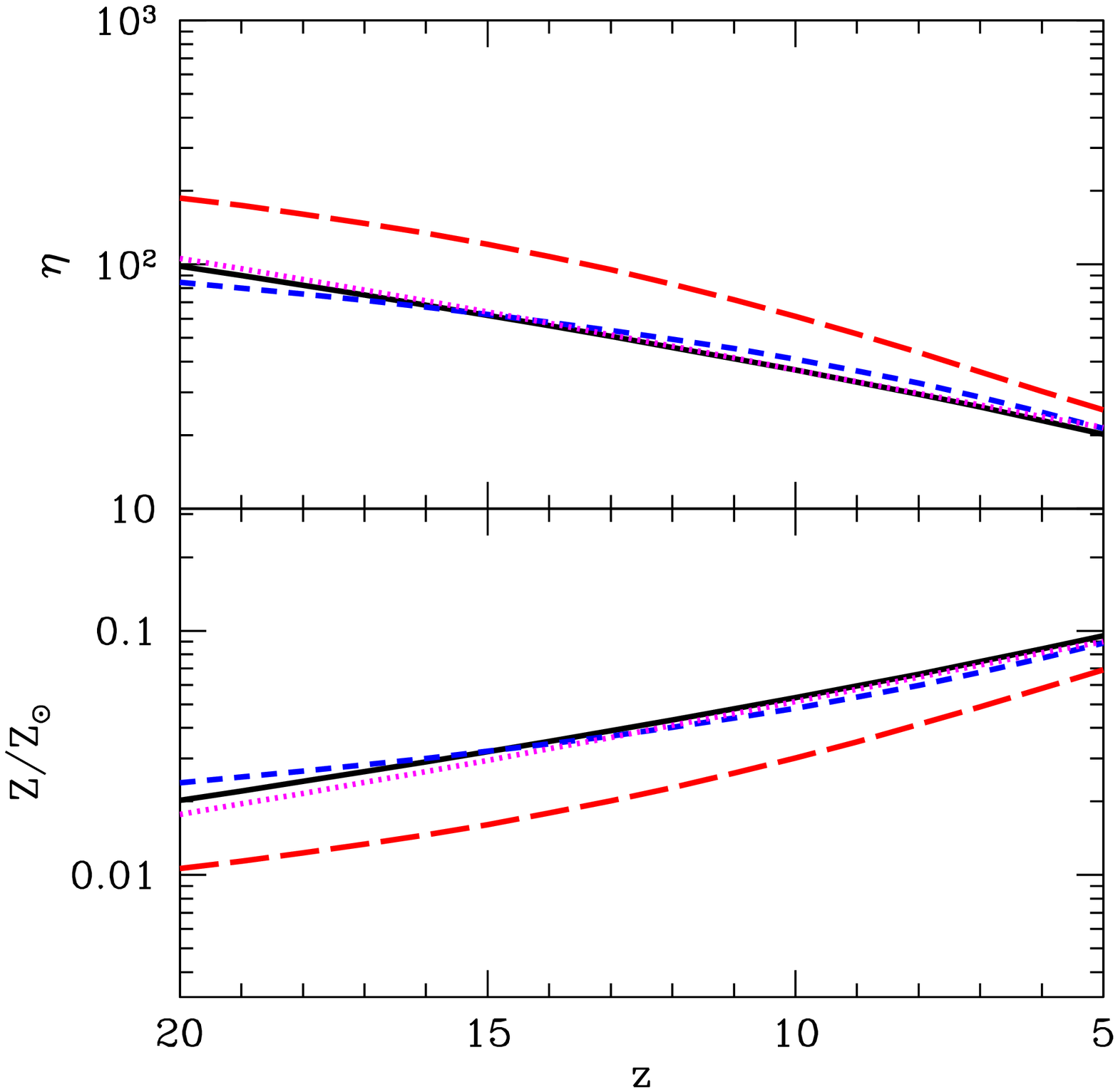} 
	\includegraphics[width=\columnwidth]{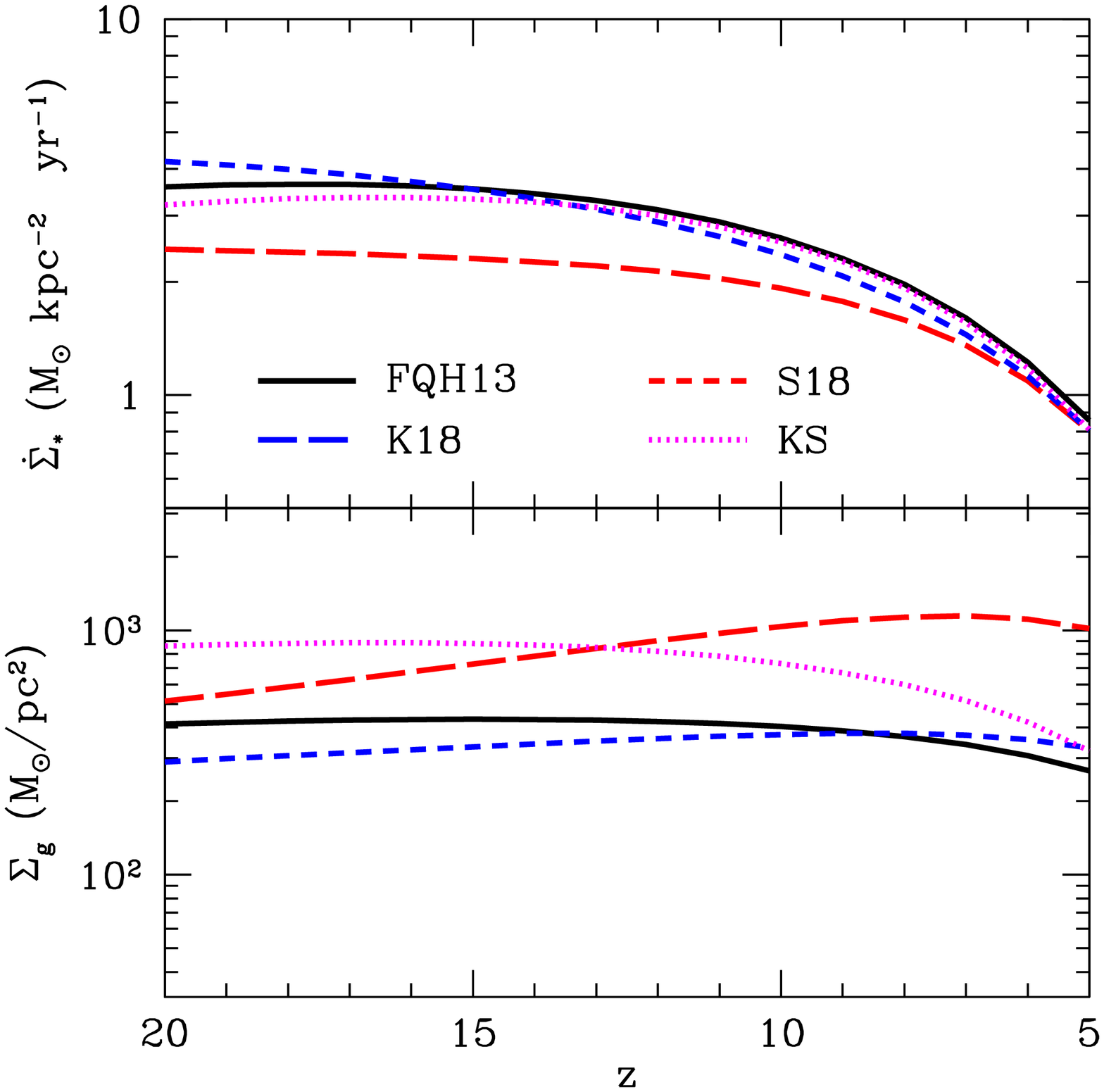} 	
    \caption{Evolution of an example galaxy with a $z=5$ mass of $10^{11} \ M_\odot$ (which begins forming stars at $z=25$). In each panel, we show results for the FQH13 model (solid curves), K18 model (long-dashed curves), S18 model (short-dashed curves), and KS law (dotted curves). \emph{Top Left:} Gas and stellar masses (thick and thin curves, respectively), compared to the total halo mass (dotted curve in bottom panel) and scaled to the baryonic mass associate with the halo (top). \emph{Top Right:} Depletion time (bottom) and specific SFR (top). Both are scaled to the Hubble time: thus the top panel is the number of stellar mass-doublings per Hubble time. \emph{Bottom Left:} Metallicity evolution (bottom) and instantaneous mass-loading parameter (top). \emph{Bottom Right:} Surface density of gas (bottom panel) and the star formation rate (top panel).}
    \label{fig:gal-sf}
\end{figure*}

We ``calibrate" the parameters of our new models (shown by the other thin curves in Figure~\ref{fig:fstar}) by ensuring that their $f_*$--$M_h$ relations are roughly similar to the momentum-regulated minimalist model. To do so, we have adjusted the feedback efficiency parameter $b_{\rm fb}$ in each case: the FQH13, K18, S18, and KS models require $b_{\rm fb} = 3.6,\, 4,\, 7,$ and 4, respectively. Given the many uncertain parameters in all of these models, and the many physical processes they ignore, we do not regard such adjustments as physically meaningful. For example, Figure~\ref{fig:bathtub-recycle} shows that assumptions about gas recycling introduce similar order unity uncertainties. In any case, unless otherwise specified we use these parameters in the remainder of the paper.

Although these adjustments work reasonably well for small haloes, they all overproduce stars when $m_h \ga 3 \times 10^{11} \, M_\odot$. There are (at least) two potential solutions to this problem invoked at lower redshifts: one is that galaxies inside massive, hot haloes do not accrete much of the gas that flows onto their halo \citep{faucher11}, while another is that the star formation is ongoing but the UV light is extinguished by dust (which should be most prevalent in these old, higher-metallicity systems). To provide a better match to the observations, we incorporate the reduced accretion rate measured in simulations by \citet{faucher11} (shown by the thick curves in Fig.~\ref{fig:fstar}), as described in detail in \citet{furl17-gal}. Dust extinction could easily make up the remaining difference \citep{mirocha20-dust}.

%%%%%%%%%%%%%%%%%%FIGURE: Basic galaxy model: star formation laws overflow (surface density)
%\begin{figure}
%	\includegraphics[width=\columnwidth]{f7.eps} 
%    \caption{Surface density of gas (bottom panel) and the star formation rate (top panel) for the same galaxy models shown in Fig.~\ref{fig:gal-sf} (which has $m_h=10^{11} \ M_\odot$ at $z=5$).  }
%    \label{fig:gal-sf-2}
%\end{figure}

\subsection{Effects of the star formation law}

Figure~\ref{fig:gal-sf} shows how an example galaxy that reaches $m_h = 10^{11} \ M_\odot$ at $z=5$ (and begins forming stars at $z_i=25$) evolves according to these four star formation prescriptions. The top left panels show the gas and stellar masses (bottom) and the retention fractions $X_g$ and $X_*$ (top; thick curves for gas and thin curves for stars). The top right panels show the gas depletion time relative to the Hubble time as well as ${\rm SSFR} \times t_H$ (i.e., the number of times the stellar mass doubles over one Hubble time). The bottom left panels show the ISM metallicity and the mass-loading parameter. The bottom right panels show the surface density of gas and of the star formation rate in the models. 

By calibrating the feedback parameters in the previous section, we have ensured that all of these models have $z=5$ stellar masses that differ by $\la 50\%$. But their stellar populations also  evolve similarly throughout their histories, as expected from the bathtub model, despite their fundamentally different approaches to star formation. The stellar masses grow exponentially, with very similar specific star formation rates that parallel the cosmological accretion rate. Additionally, within each model the gas fraction $X_g$ and the surface density $\Sigma_g$ vary only slowly over this entire time interval. The relative constancy of $X_g$ can be understood through our analytic treatment in section \ref{insight-power-law}. 

The most obvious difference between the models is in the gas mass: as in the bathtub model, the star formation rate is mostly determined by the feedback prescription, so the gas mass must adjust, according to the star formation law, to fulfill that condition. The bottom right part of Figure~\ref{fig:gal-sf} helps explain the resulting gas fractions. Most of our star formation prescriptions are based on the gas surface density. We have $\Sigma_g \sim 10^2$--$10^3 \ M_\odot$~pc$^{-2}$ -- characteristic of starbursts at lower redshifts. Our calibration of the different star formation laws essentially requires that they all have similar $\dot{\Sigma}_*$ in this regime (as seen in Figure~\ref{fig:sfr-laws}), but there are still some variations. The FQH13 model predicts very large star formation rates, because vigorous star formation is required to support such discs ($\dot{\Sigma}_* \propto \Sigma_g^2$); the K18 model is slightly more efficient in this regime, as we have normalized it. The ``required" gas mass is thus relatively small. The KS prescription, on the other hand, requires the disc to be more massive in order to generate sufficient star formation. 

%%%%%%%%%%%%%%%%%FIGURE: Basic galaxy model: Semenov parameters
\begin{figure}
	\includegraphics[width=\columnwidth]{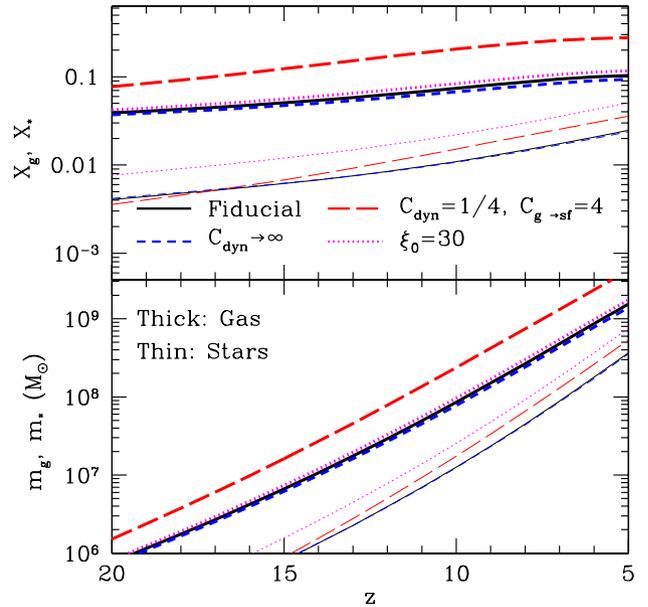} 
    \caption{Evolution of a disk galaxy (with has $m_h=10^{11} \ M_\odot$ at $z=5$) in the S18 model. The panels are similar to the left panel of Fig.~\ref{fig:gal-sf} The solid curves use our fiducial parameter choices. The others vary the parameters of the S18 model.}
    \label{fig:gal-sf-sem}
\end{figure}

%%%%%%%%%%%%%%%%%FIGURE: Basic galaxy model: feedback parameters
\begin{figure*}
	\includegraphics[width=0.66\columnwidth]{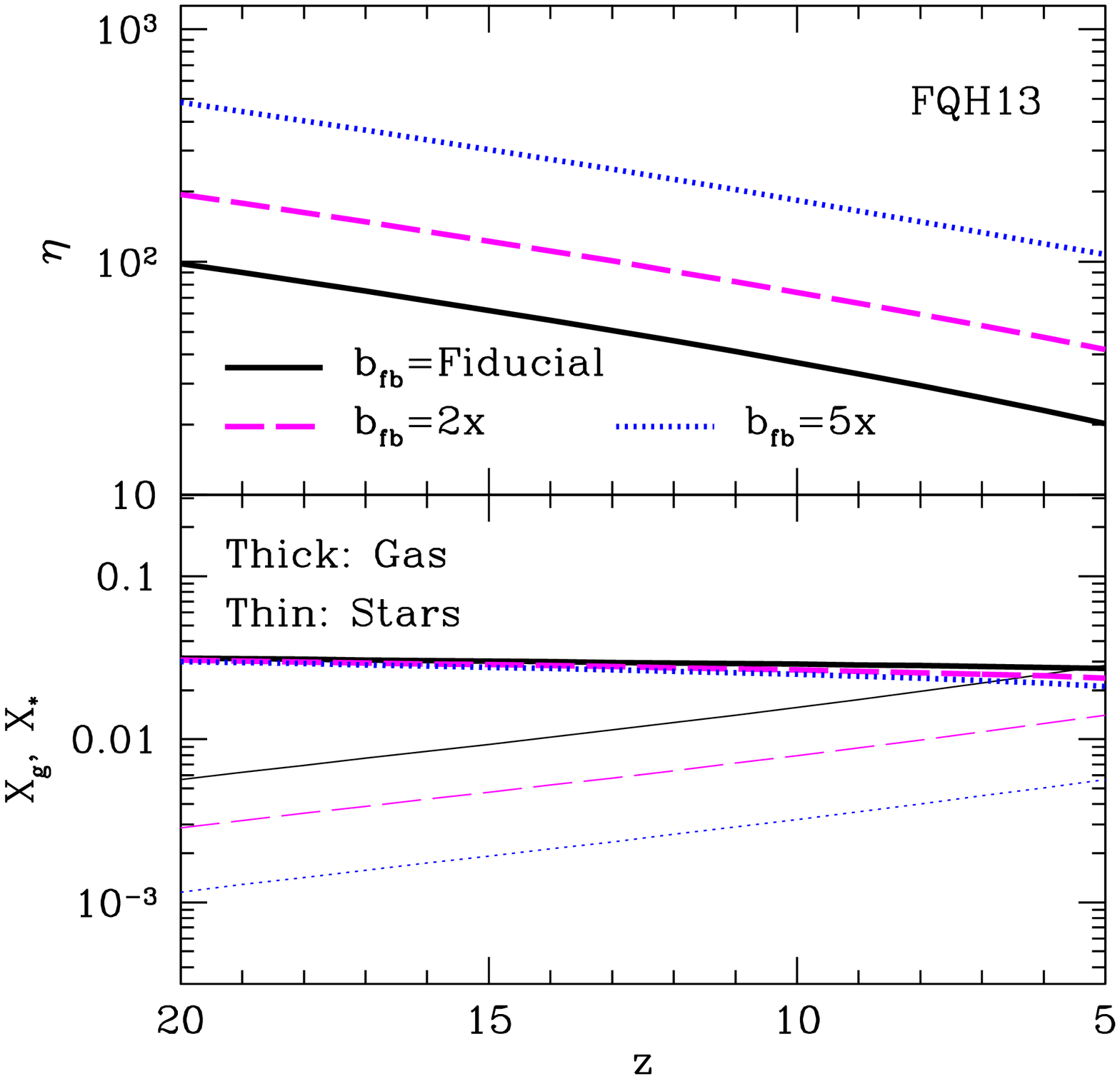} 
	\includegraphics[width=0.66\columnwidth]{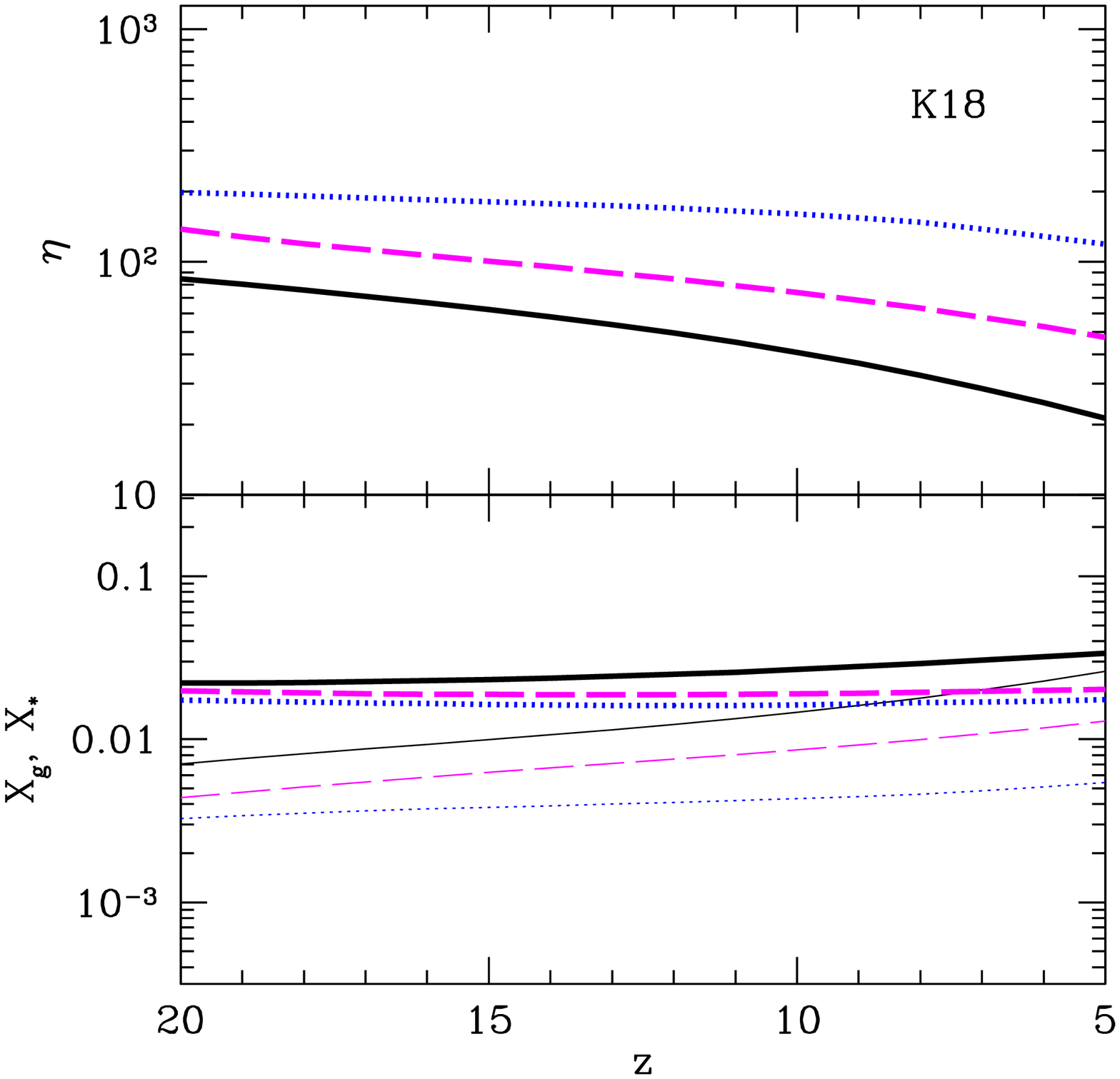} 
	\includegraphics[width=0.66\columnwidth]{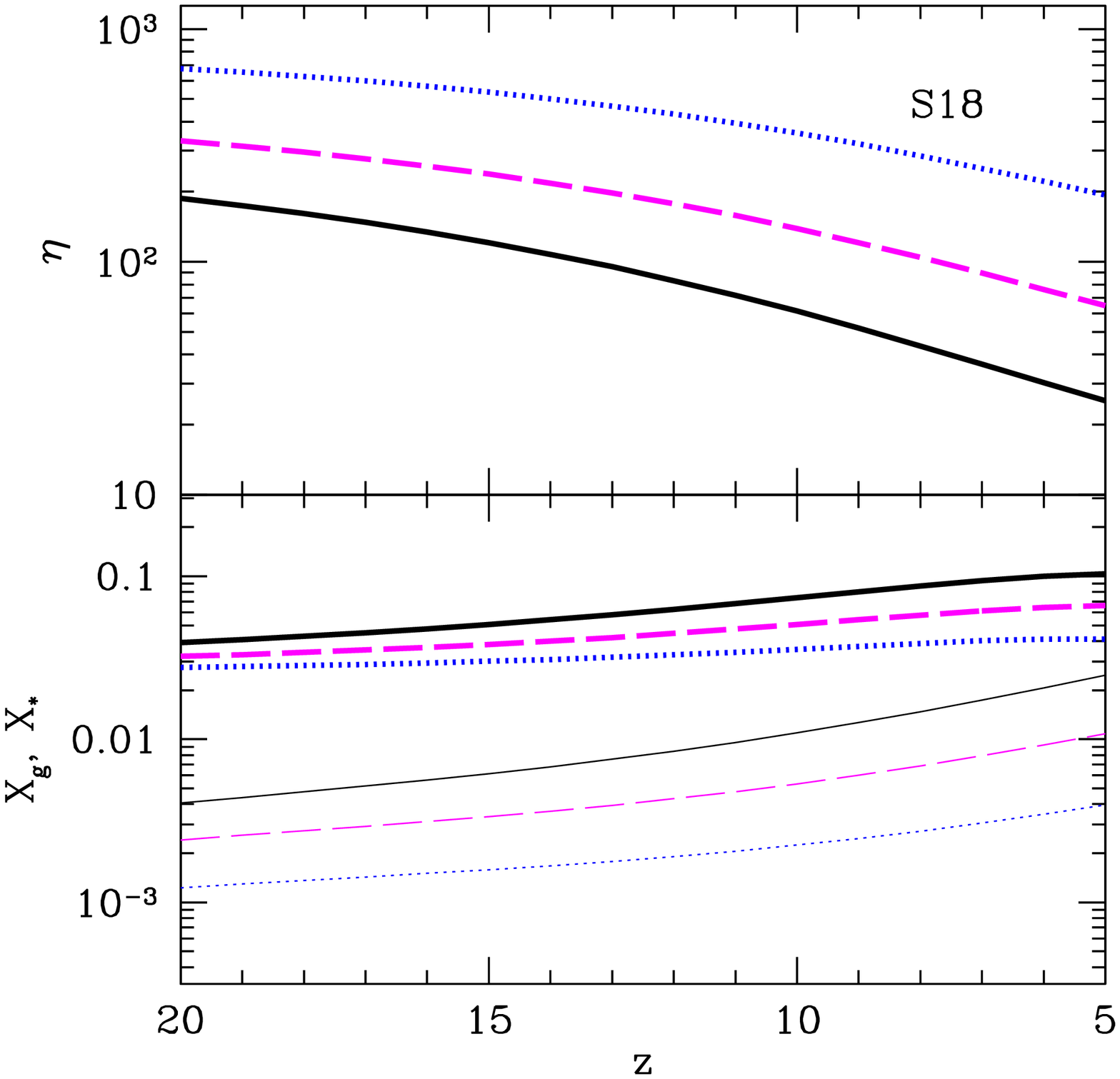} 
    \caption{The effects of feedback strength on our galaxy formation models. The columns use the FQH13 (left), K18 (center), and S18 (right) star formation models. In each panel, the solid curve uses our default feedback strength, while the dashed and dotted curves increase the strength by factors of two and five, respectively. In each column, the upper panels show the instantaneous mass-loading parameter (evaluated at each redshift) and the lower panels show the fraction of the halo's baryons in the ISM gas (thick curves) and stellar component (thin curves).}
    \label{fig:gal-feedback}
\end{figure*}

The S18 model differs from the others in that it does not rely on the gas surface density to determine the star formation rate; indeed, it makes no assumptions about how the disc is supported. Thus the star formation rate only increases linearly with the total gas mass, requiring the surface density to increase substantially.\footnote{Alternatively, the gas disc could be physically larger.} (The increased gas mass does decrease the dynamical time of the galaxy and hence the star formation timescale, but that is a very modest effect.) 

Because the star formation rate in the S18 model varies slowly with surface density, it is relatively difficult to match it to the surface-density based prescriptions.\footnote{Note that, with our assumption that $\Sigma_g = m_g/(\pi r_d^2)$ and that stars form on a dynamical time, the S18 model has $\dot{\Sigma}_* \propto \Sigma_g/t_{\rm orb}$, which is the behavior to which the K18 model also converges if $f_{\rm sf} \approx 1$. Nevertheless, the S18 prescriptions are less efficient overall.} We have already fixed the parameters of this model differently from those used by S18 to match simulations of Milky Way-type galaxies: we decreased the timescale for cloud formation and set the time over which dynamical processes destroy clouds to be significantly longer (in a relative sense). Both of these increase the star formation rate, decreasing the surface density so that it becomes closer to the other models. Figure~\ref{fig:gal-sf-sem} illustrates these effects. The solid curves use our fiducial parameters. The long-dashed curves set the timescales for cloud formation and dynamical destruction to be comparable to those found in the S18 calculations, $C_{\rm g \rightarrow sf} =4$ and $C_{\rm dyn}=0.25$. The slower rate of cloud formation, coupled with the more rapid destruction, decreases the overall star formation efficiency, substantially increasing the gas content of the discs (but also modestly increasing the overall stellar mass at late times). The dotted curves use our fiducial timescales but decrease the \emph{local} mass-loading parameter $\xi_0$ by a factor of two (without changing the global mass-loading parameter); this increases both the gas mass and the stellar mass, as each cloud transforms more of its mass into stars. Finally, the short-dashed curves ignore the dynamical destruction of clouds, illustrating that this is already unimportant with our fiducial parameters.

\subsection{Effects of feedback} \label{disc-feedback}

Figure~\ref{fig:gal-feedback} focuses on the effects of the feedback parameter, $P_*/m_*$. The solid curves take our default values for each model, while the dashed and dotted curves increase it by factors of two and five, respectively. The three columns use different star formation laws (FQH13, K18, and S18, from left to right). In the simple feedback model of section \ref{fg-simple}, the mass-loading parameter is simply proportional to this efficiency factor. In the more complex feedback model of \citet{hayward17}, the dependence is more subtle, but a linear relationship is not far off.

Recall from our discussion of the minimal bathtub model that the feedback strength affected the galaxies in two ways. First, $X_* \propto 1/\eta$, because an increased feedback efficiency evacuates more gas (i.e., fuel for star formation). This is clearly true in all the models: the stellar mass is ultimately determined by feedback. The bathtub model also predicted that $X_g \propto 1/(\eta \varepsilon)$: the gas mass adjusts to ensure that the ``correct" stellar mass is created, given a particular star formation efficiency (see Fig.~\ref{fig:bathtub-feedback}). It may therefore seem surprising that the gas retention fraction $X_g$ is very \emph{insensitive} to the strength of feedback in all of these models. This is most obvious in the FQH13 picture, and the reason is simple: that model assumes that the star formation rate is determined \emph{locally} by the need to support the disc through feedback. In that limit, we essentially have $\varepsilon \propto 1/\eta$ (c.f., eq.~\ref{eq:eps-sf-Q=1}), so that $X_g$ is constant. The other models have some sensitivity, but in both cases they still rely on stellar feedback to determine the star formation rate, so the variation of $X_g$ is much smaller than naively expected from the bathtub model.

%%%%%%%%%%%%%%%%%FIGURE: Recycling
\begin{figure}
	\includegraphics[width=\columnwidth]{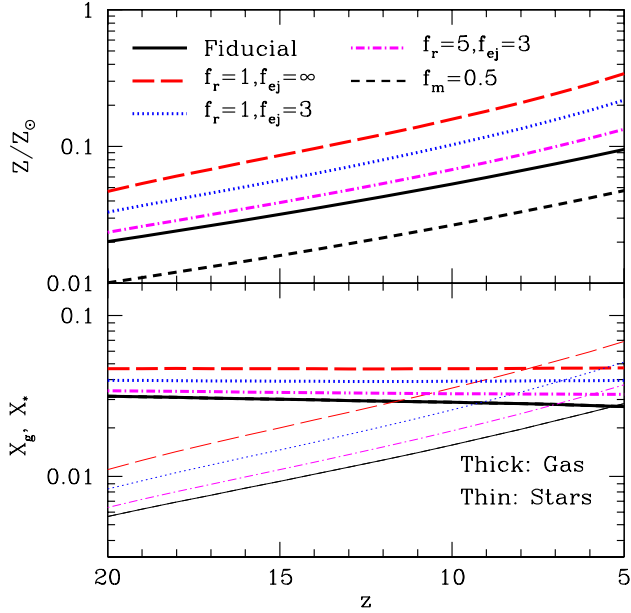} 
    \caption{The effects of gas recycling and metal mixing on our galaxy models. The solid curves show our fiducial model with the FQH13 star formation law, without gas recycling and assuming perfect metal mixing ($f_m=0$). The short-dashed curve in the top panel shows the metallicity if $f_m=0.5$ of the metals are ejected immediately in the outflow. The other curves vary the gas recycling parameters. }
    \label{fig:gal-recycle}
\end{figure}

\subsection{Gas recycling}

So far in this section, we have assumed that all the gas ejected from the disc escapes the system entirely. In Figure~\ref{fig:gal-recycle}, we re-introduce the gas reservoir near the galaxy and allow the material to re-accrete onto the disc, as described in section \ref{reservoir}. The solid curves are identical to those from Figure~\ref{fig:gal-sf}, ignoring recycling. The long-dashed curves assume that all of the gas remains bound to the galaxy's halo and can re-accrete on a timescale $t_{\rm dyn}$ (i.e., $f_r=1$). This substantially increases the stellar mass and metallicity of the galaxy, with a somewhat smaller effect on the gas mass. The dotted and dot-dashed curves decrease the importance of recycling, by allowing some gas to escape the halo entirely and/or by increasing the re-accretion timescale. Importantly, however, the addition of recycling has only a modest qualitative effect on our models -- it certainly increases the star formation rate, but it does not affect the overall exponential growth, the specific star formation rate, or the gas fractions.

%%%%%%%%%%%%%%%%%FIGURE: Mass trends: star formation laws
\begin{figure*}
	\includegraphics[width=0.66\columnwidth]{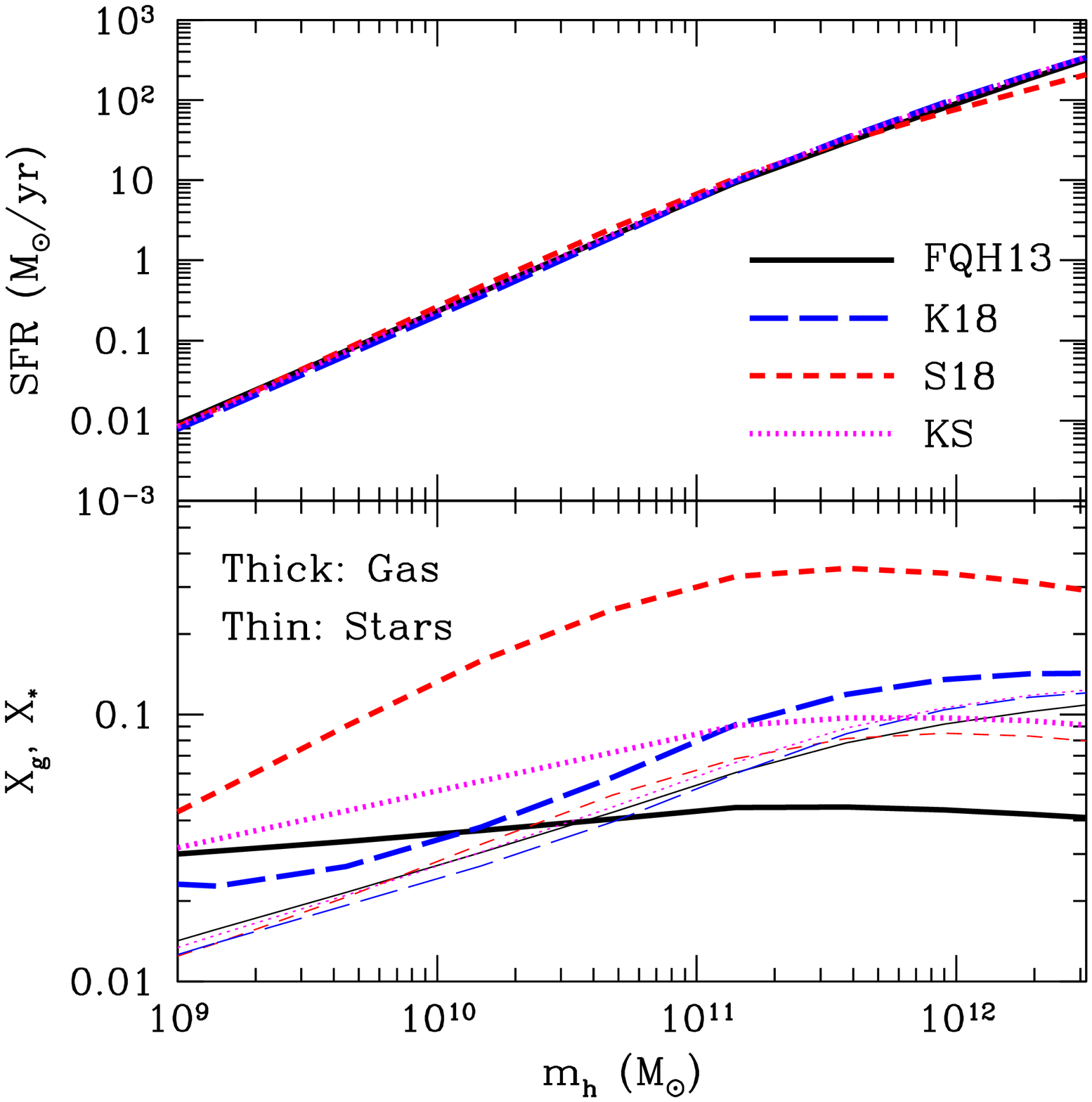} 
	\includegraphics[width=0.66\columnwidth]{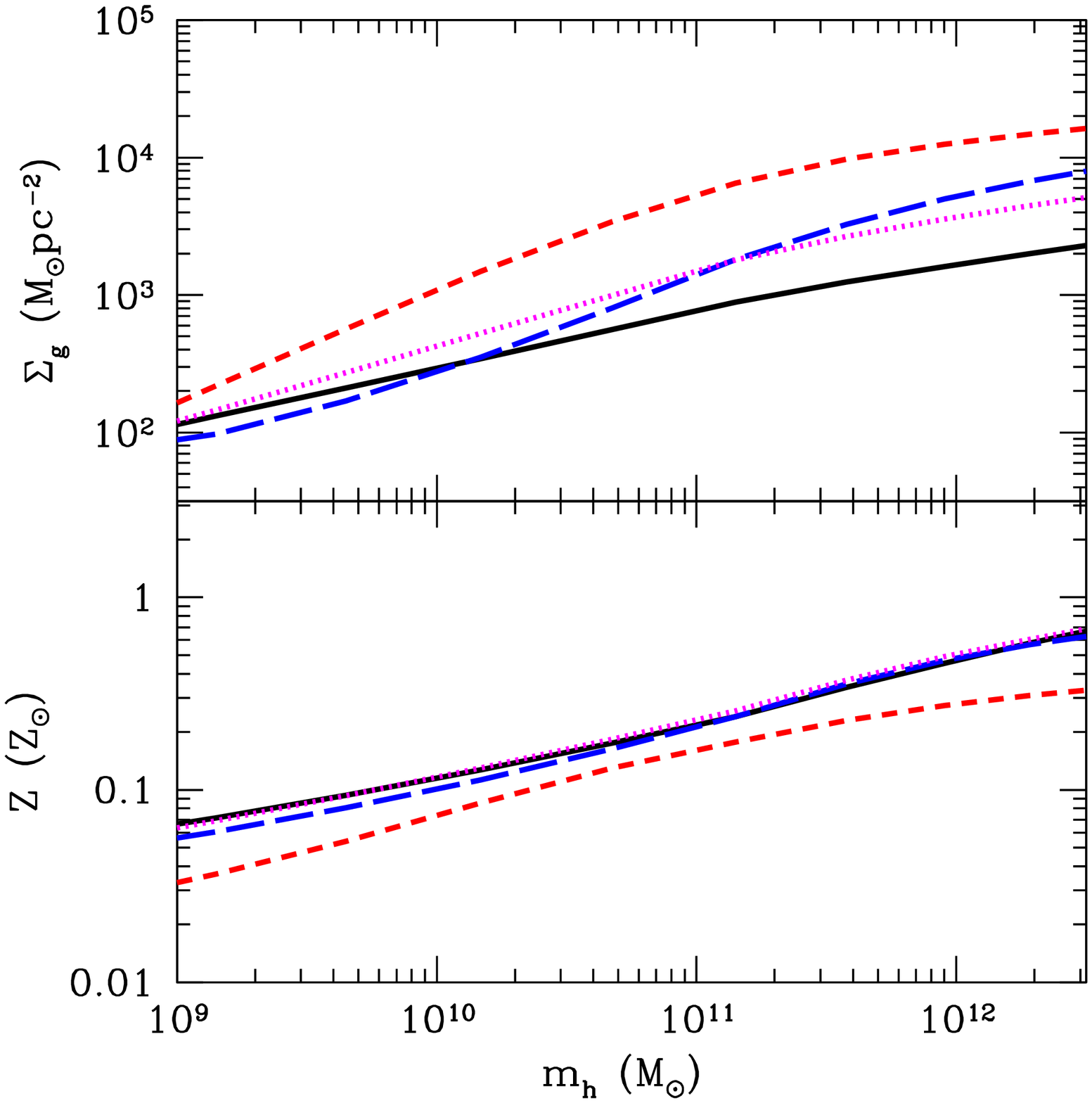} 
	\includegraphics[width=0.66\columnwidth]{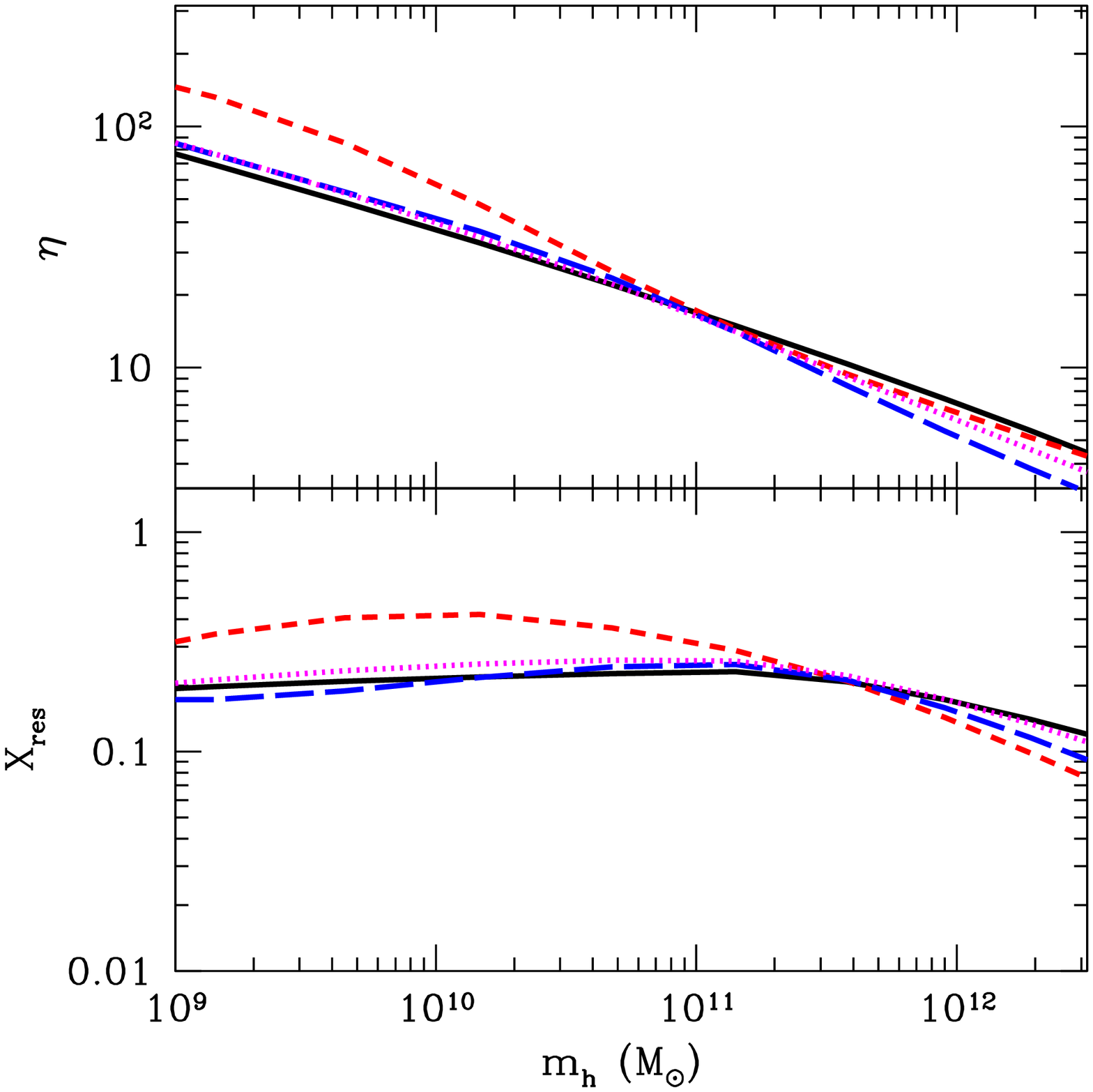} 
    \caption{Galaxy properties at $z=7$ in several of our models, as a function of halo mass. \emph{Left:} The star formation rates (top) and stellar and gas retention fractions (bottom; thin and thick curves, respectively). \emph{Center:} Gas surface density (top) and metallicity (bottom). \emph{Right:} Instantaneous mass-loading factors (top) and retention fraction for gas ejected from the disc but still bound to the halo, $X_{\rm res}$ (bottom). }
    \label{fig:galmass}
\end{figure*}

\subsection{Chemical evolution}

The bottom left panel of Figure~\ref{fig:gal-sf} shows the metallicity evolution for our four different star formation laws. The metallicity increases exponentially with $z$, reaching $Z \sim 0.1 Z_\odot$ when the halo reaches $m_h \approx 10^{11} \ M_\odot$. This is easy to understand: both the stellar mass (to which the metal output is proportional) and gas mass increase exponentially with time, and because $Z=m_Z/m_g$, the metallicity will grow exponentially at a rate determined by the difference between the stellar mass timescale and the gas disc growth timescale. The metallicity is quite small for most of the galaxy's evolution, simply because the gas disk is so much more massive than the stellar phase.  

Additionally, note that even at $z=5$ our model galaxy has a metallicity significantly smaller than comparable systems in the local Universe. For example, \citet{tremonti04}, who measured the mass-metallicity relation at $z \sim 0$, found $Z \sim 0.5 \ Z_\odot$ for $m_* \sim 10^{8.5} \ M_\odot$ (the stellar mass of this galaxy). However, our results are much closer to galaxies in the star-forming era. For example, \citet{sanders20} found $Z \sim 0.2 \ Z_\odot$ at $z \sim 3$, if we extrapolate their fit (which as made at $m_* > 10^9 \ M_\odot$) to the same stellar mass. The normalization of our model results from a complex interplay of accreting gas (which we assume to be primordial) and feedback, the latter of which depends somewhat on the star formation model. For example, at early times the KS model has a slightly smaller metallicity than FQH13, because it requires a larger surface density.  The S18 model has the lowest metallicity, even though it also has the highest gas mass.

The normalization also depends on assumptions about metal mixing. For example, the short-dashed curves in Figure~\ref{fig:gal-recycle} set $f_m=0.5$, so that half of the metals are immediately ejected from the galaxy. Unsurprisingly, this reduces the ISM metallicity by that same factor. This decrease can be moderated if some of the ejected gas is allowed to re-accrete: the other curves show how gas recycling increases the metallicity, by bringing enriched gas back into the system.

\section{Galaxy populations} \label{mass-trends}

Having examined how individual galaxies evolve over time, we now consider how their properties depend on halo mass at a fixed redshift. Henceforth, we will include gas recycling (with $f_r=1$ and $f_{\rm ej}=3$) as well as the suppression of accretion at high masses described in section \ref{lf-fit}. Otherwise, our parameters will be identical to the previous section. 

\subsection{Trends with halo mass}

Figure~\ref{fig:galmass} shows several properties of our galaxies at $z=7$ (we will consider other redshifts below). The left column shows the star formation rate and gas/stellar retention fractions (thick and thin curves, respectively), the middle column shows the gas surface density and metallicity, and the right column shows the instantaneous mass-loading factor and the fraction of baryonic mass associated with the halo that is in the gas reservoir, $X_{\rm res}$ (i.e., ejected from the disc but still bound to the halo). 

Several important trends are obvious from Figure~\ref{fig:galmass}. First, all the prescriptions provide very similar star formation rates across the entire mass range The star formation rate increases nearly as $m^{4/3}$, which simply reflects the mass dependence of momentum-regulated feedback. The upper right panel further illustrates this point: the FQH13 model has $\eta \propto m^{-1/3}$, though the other models have slightly steeper declines.

\begin{figure}
	\includegraphics[width=\columnwidth]{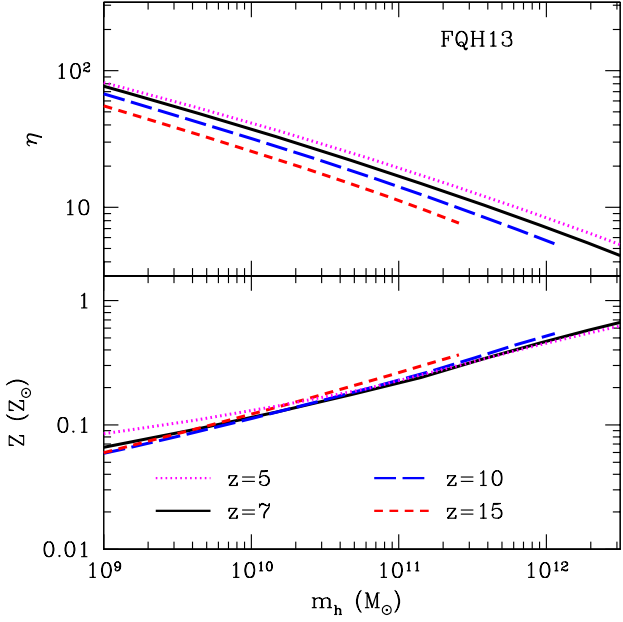} 
    \caption{Redshift evolution of galaxy properties in the FQH13 model. The upper panel shows the instantaneous mass-loading factor, while the lower panel shows the gas phase metallicity. The other star formation prescriptions  have similar redshift dependence. }
    \label{fig:galmass-z-feedback}
\end{figure}

The upper panel in Figure~\ref{fig:galmass-z-feedback} examines the redshift evolution of $\eta$. Here we use the FQH13 star formation model, but the other models have very similar redshift dependence: we find that $\eta \propto (1+z)^{1/2}$, as expected for a naive implementation of momentum-regulated feedback. This scaling causes tension with observations \citep{mirocha20} and may require additional physics to match measurements.

Most of the models have discs that retain an increasing fraction of their gas with halo mass (at least until $m_h \sim 10^{11.5} \ M_\odot$, where the suppression of accretion becomes important), but the FQH13 model is an exception. In this model, we have already seen that because the effective star formation efficiency in the disc is inversely proportional to the feedback strength, $X_g$ is independent of that value. Because the feedback is the only part of this model that depends on the mass, $X_g\approx$constant in that case. In the other models, these factors do not cancel, so more massive halos have relatively larger gas discs, although  never by more than about an order of magnitude. In the S18 case, the discs retain $\ga 10\%$ of the halo's baryonic mass over a broad range of halo masses, approaching an order of magnitude more than the FQH13 discs. This further demonstrates that the gas masses of galaxy discs can provide a useful probe of the mechanisms driving star formation, although we note that the disc masses are certainly also sensitive to other aspects of the model (such as gas recycling).

The lower center panel in Figure~\ref{fig:galmass} shows that the metallicity increases with halo mass, as one naively expects. All of our models have roughly $Z \propto m_h^{1/3}$; we will compare this relation to observations in section \ref{trend-stellar-mass} below. Figure~\ref{fig:galmass-z-feedback} also shows that the metallicity at a fixed halo mass depends only weakly on redshift, with slightly higher values at earlier times, because the halos are more tightly bound at higher redshifts so produce more stars (but also retain more gas in their discs). 

The lower right panel of Figure~\ref{fig:galmass} shows that, in all of the models, a substantial fraction of the gas remains in the gas reservoir, even though we have allowed gas ejected at high velocity to escape the halo and chosen a relatively fast recycling timescale for these models. This reservoir is a very crude representation of the circumgalactic medium, and our models show that such a medium will likely remain important in high-$z$ galaxies. However, note that its properties are not particularly sensitive to the star formation law.

%%%%%%%%%%%%%%%%%FIGURE: Mass trends: redshift feedback

\begin{figure}
	\includegraphics[width=\columnwidth]{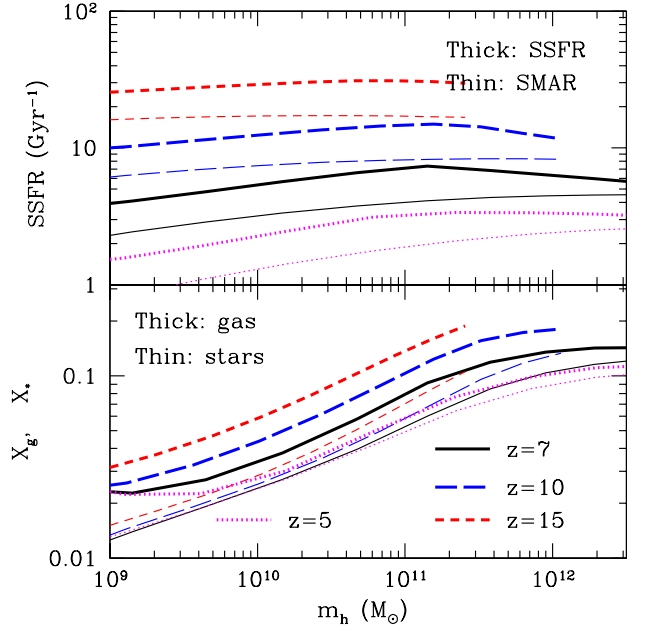} 
    \caption{Redshift evolution of galaxy properties in the K18 model. The upper panel shows the specific star formation rate (thick curves) and specific mass accretion rate (thin curves), while the lower panel shows the gas and stellar retention fractions.  }
    \label{fig:galmass-z}
\end{figure}

In Figure~\ref{fig:galmass-z} we show how the gas and stellar retention fractions and specific star formation rates depend on redshift; here we use the K18 model. Note that the SSFR depends only slightly on halo mass and is always a multiple of the specific mass accretion rate, which increases rapidly with redshift. The gas disc mass also increases with redshift; this is true of all the  models, though the redshift dependence is weaker in the FQH13 model. The stellar mass fraction, on the other hand, changes only slowly with redshift in the K18 model, but the dependence is stronger in other models. 

\subsection{Trends with stellar mass} \label{trend-stellar-mass}

\begin{figure}
	\includegraphics[width=\columnwidth]{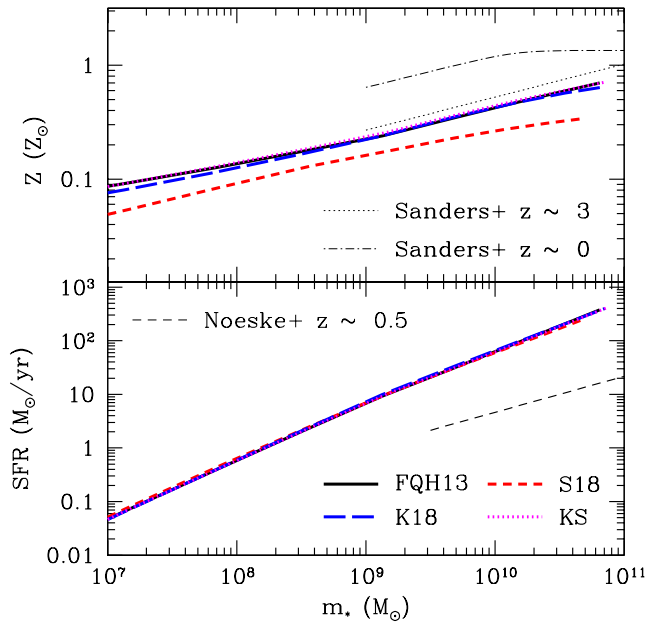} 
    \caption{The mass-metallicity relation (top panel) and star formation rate (bottom panel) as a function of stellar mass at $z=7$. In the top panel, we show measurements from \citet{sanders20} at $z \sim 0$ and $3$ with the thin dot-dashed and dotted curves, respectively. In the lower panel, we show the ``star-forming main sequence" at $z \sim 0.5$ \citep{noeske07} with the thin dotted curve.}
    \label{fig:sf-ms}
\end{figure}

Although variations in galaxy properties with halo mass are useful from a theoretical perspective, the stellar mass is a much more straightforward observable (though still very difficult to measure at such high redshifts!). To that end, Figure~\ref{fig:sf-ms} shows how two important quantities vary with stellar mass at $z=7$: the mass-metallicity relation (top panel) and the star formation rate (bottom panel). In both panels the solid, long-dashed, short-dashed, and dotted curves show the FQH13, K18, S18, and KS star formation prescriptions. 

%%%%%%%%%%%%%%%%%FIGURE: Mass trends: redshift

In the bottom panel, we see that the star formation rate increases nearly linearly with stellar mass, aside from a slight turnover at large masses because of the suppression of accretion in this regime. This is just another way of saying that the specific star formation rate is also very nearly constant across the galaxy population (within a given model, at least). All the models are nearly the same, because we have chosen parameters that roughly match each other. The thin dotted line shows the measured SFR-$m_*$ relation along the ``star-forming main sequence" at $z\sim 0.5$ from \citet{noeske07}, who measured the relation at $m_* \ga 10^{9.5} \ M_\odot$, so we can only compare to the most massive of our galaxies. Unsurprisingly, our models have much more rapid star formation at a fixed stellar mass; this is because accretion is so much faster at high redshifts. More interestingly, the $z \sim 0.5$ relation is sublinear and hence shallower than our model. Of course, it is not clear how meaningful the comparison to local results is, given the very different cosmological context of rapidly-accreting high-$z$ galaxies.

All of our models have similar metallicity trends. The metallicity increases with stellar mass as $Z \propto m_*^{0.25-0.3}$. This slope can be understood quite simply. The metallicity scales as $Z \propto m_*/m_g$. With momentum-regulated feedback, the star formation efficiency is $\propto \eta^{-1} \propto m_h^{1/3}$, so $m_* \propto m_h^{4/3}$. We have found that the gas retention fraction $X_g$ is roughly independent of mass in all of our models, so $m_g \propto m_h$. In this very rough estimate, then, $Z \propto m_h^{1/3} \propto m_*^{1/4}$, which is close to the relation we find in the full models. For context, we compare the models to the observed relations of \citet{sanders20} at $z \sim 0$ and $3$. The local relation has a fairly complex shape, but the $z \sim 3$ relation has $Z \propto m_*^{0.29}$, which is very close to the models. Interestingly, the normalization is also close to the $z \sim 3$ value (though far from the local relation). 

We have seen in Figure~\ref{fig:galmass-z-feedback} that our models predict very little redshift dependence to the halo mass-metallicity relation. Although we do not show it here, we note that in our models the mass-metallicity relation is nearly independent of redshift. This is because both the gas mass and stellar mass depend on feedback in similar ways in feedback-driven models (see the discussion in section~\ref{disc-feedback}), so that the ratio between the two is nearly constant. This is interesting in light of the similarity of our model results to the \citet{sanders20} observations at $z \sim 3$, although they were only able to measure this slope in galaxies with $m_\star \ga 10^9 \ M_\odot$.. While we cannot extend our results all the way to $z \sim 3$, it is reassuring that they are comparable, since both focus on star-forming galaxies.

%%%%%%%%%%%%%%%%%FIGURE: Trends with stellar mass

\section{Discussion} \label{disc}

A deep understanding of galaxy formation and evolution is, without question, a fundamental goal of astrophysics. The past decade has seen a revolution in our understanding of many facets of galaxy evolution, using a wide range of models -- from the simplest analytic approaches to detailed numerical simulations. While all of these approaches have provided important insights, it is challenging to compare the general conclusions of simple models with more detailed calculations in a way that improves our physical insight. In this paper, we have extended a class of ``minimalist" models to include more sophisticated prescriptions for star formation and feedback, thought to be the most fundamental processes driving galaxy evolution. While analytic, the deeper models are inspired by the results of numerical simulations. We have included star formation models that rely on stellar feedback to support the disc (FQH13), include molecular gas (K18), and treat star formation as regulated by gas cycling through the phases of the ISM (S18). Additionally, we include feedback in a turbulent ISM \citep{thompson16, hayward17}. We have focused on high-$z$ galaxies largely because we can ignore many of the more complicated aspects of galaxy formation, such as ``quenching," in this era. 

We found that most of the conclusions of the simplest models remain valid at this increased level of complexity. In particular: \emph{(i)} the overall star formation rate in a galaxy is determined by the feedback efficiency (in our model, the mass-loading factor) and is nearly independent of the details of star formation; \emph{(ii)} the specific star formation rate is nearly independent of galactic processes and depends almost entirely on the cosmological accretion rate, so that the stellar mass of a galaxy increases exponentially with redshift; \emph{(iii)} the ISM mass also increases exponentially with redshift, maintaining a nearly constant fraction of the halo mass, with its normalization determined by the star formation law. This means that the the ISM mass ``self-adjusts" so that the star formation rate reaches the level required by the feedback law; however, in models in which feedback regulates star formation on a \emph{local} level, the ISM mass is much less sensitive to feedback than expected from the minimalist bathtub model. 

Additionally, we found that our more sophisticated models obey relatively simple trends across halo mass and redshift that can be well-approximated by the sorts of prescriptions used in semi-empirical or ``minimalist" analytic models of galaxy formation. For example, the star formation efficiency is roughly a power law (or broken power law) with halo mass, and the feedback mass-loading parameter obeys the expected mass and redshift trends from momentum-regulated feedback. This is heartening, because it means that such simple laws -- useful in that they can be easily incorporated into state-of-the-art fitting procedures -- can be directly connected to physical parameters of galaxies, even if the latter are built from more detailed physics.

Incorporating more sophisticated physics into analytic models of galaxy formation offers the additional benefit of helping to identify observables that will enable us to measure such physical processes. For example, our models all predict that the star formation rate is nearly independent of the ``microphysics" of star formation but depends directly on the feedback efficiency. Only with simultaneous measurements of the ISM gas can the star formation law itself be probed. We also find that the gas-phase metallicity increases steadily with halo mass, as one would expect from population studies at later cosmic times, largely because smaller halos have had less time to form stars.

The necessity of probing the gas content in order to understand star formation processes at these early times emphasizes the importance of multiwavelength observations of high-$z$ galaxies. While ALMA has already measured some of these properties through dust and the [CII] and [OIII] lines (e.g., \citealt{smit18, carniani18, lefevre19}), this is generally only possible for relatively bright sources. Probing earlier phases in galaxy formation, when the haloes are still small, may be easiest with intensity mapping, in which coarse resolution is used to study fluctuations in the integrated galaxy populations \citep{kovetz19, chang19}. Intensity mapping is powerful because the cumulative light is dominated by faint sources at these early redshifts (e.g., \citealt{robertson15}); however, metal emission lines are only indirect probes of the ISM mass, so their interpretation will certainly require more detailed modeling \citep{ferrara19, yang20}.  

Of course, our models are far from complete, and any number of processes that we have neglected could compromise our conclusions. In that sense, they provide a baseline set of models for a ``null hypothesis" test of galaxy physics at early times. Because our models have shown that qualitatively different star formation prescriptions make little difference to the galaxy histories, more detailed descriptions of star formation seem unlikely to change our conclusions in a qualitative sense. But we have made a number of strong assumptions about galaxy evolution that will break down in certain populations. For example, most of our models are predicated on a ``quasi-equilibrium" balance between feedback and star formation. In the early phases of galaxy evolution, when the dynamical time is very short, such equilibrium may be difficult to establish \citep{faucher18}. The resulting burstiness certainly affects our parameterization of disc support through stellar feedback, though it is not clear how. The equilibrium assumption also does not appear to work in local dwarf galaxies (because their star formation timescales are too long), suggesting more caution in applying it to small systems \citep{forbes14b}. We have also ignored mergers in our galaxy evolution models, which will contribute to burstiness as well. 

We must also note that, although our physical models can be approximated by simple analytic models, that does not mean that such models provide an accurate representation of galaxy observations. For example, semi-empirical and minimalist models have found that the best matches to observed luminosity functions are provided by models in which the star formation efficiency is independent of redshift. Our models do predict redshift evolution in the feedback efficiency (and hence star formation efficiency), because the binding energy of halos increases at higher redshifts. It is not clear whether this disparity points to a fundamental flaw of feedback models, but it certainly deserves further study (e.g., \citealt{mirocha20}).

In this paper, we have not attempted to fit these models to observations. Measurements of luminosity functions allow us to calibrate the models, and complementary observations of high-$z$ galaxies, including spectra, clustering, and ISM properties, that are now becoming possible will probe galaxy evolution in detail during this era. We will explore how such observations can probe our models in future work.

\section*{Acknowledgements}

We thank J.~Mirocha and A.~Trapp for very helpful conversations and comments and the anonymous referee for comments that improved the manuscript. This work was supported by the National Science Foundation through award AST-1812458. In addition, this work was directly supported by the NASA Solar System Exploration Research Virtual Institute cooperative agreement number 80ARC017M0006. We also acknowledge a NASA contract supporting the ``WFIRST Extragalactic Potential Observations (EXPO) Science Investigation Team" (15-WFIRST15-0004), administered by GSFC. This material is based upon work supported by the National Science Foundation under Grant Nos. 1636646 and 1836019, the Gordon and Betty Moore Foundation, and institutional support from the HERA collaboration partners.

\section*{Data Availability}

No new data were generated or analysed in support of this research.

%%%%%%%%%%%%%%%%%%%%%%%%%%%%%%%%%%%%%%%%%%%%%%%%%%



\bibliographystyle{mnras}
\bibliography{Ref_composite} % if your bibtex file is called example.bib

% Don't change these lines
\bsp	% typesetting comment
\label{lastpage}
\end{document}